\documentclass[aps,prl,superscriptaddress,groupedaddress]{revtex4}  
\usepackage{graphicx}   
\usepackage{dcolumn}    
\usepackage{bm}  
\usepackage{tabularx}      
\usepackage{amssymb}   
\usepackage{amsmath}
\usepackage[T1]{fontenc}
\usepackage{lineno}
\usepackage{xcolor}
\begin{document}
	\title{Geometrically Regulating Evolutionary Dynamics in Biofilms}
	\author{Youness Azimzade}
	\email[Email:~]{younessazimzade@gmail.com}
	\affiliation{Department of Physics, University of Tehran, Tehran 14395-547, Iran}
	\author{Abbas Ali Saberi}
	\email[Email:~]{ab.saberi@ut.ac.ir}
	\affiliation{Department of Physics, University of Tehran, Tehran 14395-547, Iran}
	\affiliation{Institut f\"ur Theoretische Physik, Universitat zu K\"oln, Z\"ulpicher Strasse 77, 50937 K\"oln, Germany}
	\date{\today}
	
	\begin{abstract}  
		Theoretical understanding of evolutionary dynamics in spatially structured populations often relies on non-spatial models. Biofilms are among such populations where a more accurate understanding is of theoretical interest and can reveal new solutions to existing challenges. Here, we studied how the geometry of the environment affects the evolutionary dynamics of expanding populations, using the Eden model. Our results show that fluctuations of sub-populations during range expansion in 2D and 3D environments are not Brownian. Furthermore, we found that the substrate's geometry interferes with the evolutionary dynamics of populations that grow upon it. Inspired by these findings, we propose a periodically wedged pattern on surfaces prone to develop biofilms.  On such patterned surfaces, natural selection becomes less effective and beneficial mutants would have a harder time establishing. Additionally, this modification accelerates genetic drift and leads to less diverse biofilms. Both interventions are highly desired for biofilms. 
	\end{abstract} 
	\maketitle  
	
	\section{Introduction.} 
	Dense communities of bacteria, known as biofilms, appear in a wide range of natural and hand-made environments and their eradication is of great concern in medical and industrial settings \cite{nadell2016spatial, yanni2019drivers}. Biofilms are comprised of different bacteria types with different properties. Due to such diversity, biofilms can develop drug resistance \cite{fitzgerald2019bacterial, sharma2019antibiotics} and/or more types with higher fitness advantages may become dominant \cite{steenackers2016experimental}, which makes their suppression even harder. Understanding determinant processes of evolutionary dynamics of biofilms and interfering with it to have the desired outcome, for example, a community with lower drug resistance capability, has been the main concern for years \cite{mira2015rational, bittihn2017suppression, fitzgerald2019bacterial, santos2019evolutionary, shibasaki2020controlling, sanchez2020directed}.
	
	Competition between different bacteria types in biofilms leads to spatial segregation  \cite{hallatschek2007genetic, oldewurtel2015differential, gralka2019environmental}. Once segregated, bacteria face their own kind except for specific areas (i.e., the domain walls), making the spatial structure even more determinant. These domain walls separate different types and play a crucial role in the evolutionary dynamics of biofilms \cite{hallatschek2007genetic, gralka2019environmental}. Due to such a role, quantitative understanding of domain wall structures has gained increasing attention in the past few years \cite{wang2016probing, hartmann2019emergence, azimzade2019short, paula2020dynamics}. Based on such studies, we know that the geometry of individuals \cite{smith2017cell} alongside ecological processes and underlying physical interactions \cite{oldewurtel2015differential, flynn2016evolution, warren2019spatiotemporal} regulate the geometry of domain walls and evolutionary dynamics. While a possible manipulation in biofilms' evolutionary dynamics is of great importance \cite{mira2015rational, santos2019evolutionary, enriquez2015application, srey2013biofilm, lutey2018towards}, all these features are intrinsic to cells and are hard, if not impossible, to alter directly. In this regard, unraveling additional parameters that contribute to the evolutionary dynamics of biofilms is of theoretical interest and may provide a window for interventions. 
	
	The spatial structure of populations can have a determinant effect on evolutionary dynamics of different populations \cite{durrett1994importance, komarova2006spatial, wodarz2020mutant}.  Various aspects of such a role have been studied for expanding populations \cite{lavrentovich2016spatially,beller2018evolution, kayser2018emergence, mobius2019collective}. Based on these results, natural selection can happen at a faster rate in expanding populations \cite{gralka2016allele}. On the other hand, the location in which a mutation appears plays a central role in its future success \cite{fusco2016excess, lamprecht2017multicolor}.  More interestingly, the structure of the environment not only affects the geometry of invasion front \cite{azimzade2019effect, azimzade2020invasion}, it can interfere with evolutionary dynamics of populations by randomly blocking beneficial mutants \cite{gralka2019environmental, mobius2015obstacles}. Its surface structure has also been studied among different environment features, revealing that the initial geometry can affect the fluctuations of subpopulations \cite{derrida1991interface, chu2019evolution}. The question we will be tackling in this work is how different properties of the surface's geometry can affect growing populations' evolutionary dynamics and how we can exploit this geometry to interfere with the evolutionary dynamics.
	
	In this letter, we studied the geometry of domain walls under different assumptions for the environment's structure. Our findings revealed that the environment's initial geometry (for example, the surface's geometry upon which biofilms grow) interferes with domain wall fluctuations. Inspired by such observation, we proposed a macro fabrication of surfaces prone to develop biofilms to regulate evolutionary dynamics. Such a modification in geometry decreases the biodiversity of two identical populations and slows down the domination of more types with higher fitness advantages during expansion. This interference with evolutionary dynamics, especially when the nature of growing populations is not clear, is of potential application in medicine and industry.
	
	\section{Model.} To implement a spatially explicit model for population dynamics that captures complex spatiotemporal interactions, we use the two species Eden model \cite{eden1961two, saito1995critical}. This model has been used in a wide variety of research on population dynamics \cite{reiter2014range, lavrentovich2014asymmetric, chu2019evolution, azimzade2019short}, primarily due to an accurate representation of bacteria range expansion \cite{saito1995critical, hallatschek2007genetic}.
	We consider two population kinds, (A) and (B), living in a (2+1)D environment of sizes $w \times l \times l$ where populations can live and expand in $z$ direction (see FIG. \ref{FIG1} (a)). $w=1$ is the traditional (1+1)D version that has been used widely in the literature \cite{horowitz2019bacterial, chu2019evolution}.
	As an initial condition, we set the left/right side of the first plain (i.e., for $x<0$ and $x>0$ at $t=0$) to be occupied by (B)/(A) and boundary condition in $x$ ($y$) direction is reflective (periodic). We randomly select lattice units and if the unit is occupied by (A)/(B), one of the possible empty nearest neighbors will be occupied by a newborn (A)/(B) with probability $R_{A(B)}$. Respectively, the duplication process follows a Binomial distribution. For considerably small values of $R_{A(B)}$, this Binomial distribution converges to Poisson distribution \cite{simons1971convergence} ---that is standard distribution in population dynamics studies. $R_{A}=R_{B}$ represents two sub-populations with identical duplication rates ($R_{A}\ne R_{B}$ which represents sub-populations with different fitness advantages will be included in the future). 
	
	\section{Results.}
\begin{figure}[h]
		\centering
		\includegraphics[width=0.42\linewidth]{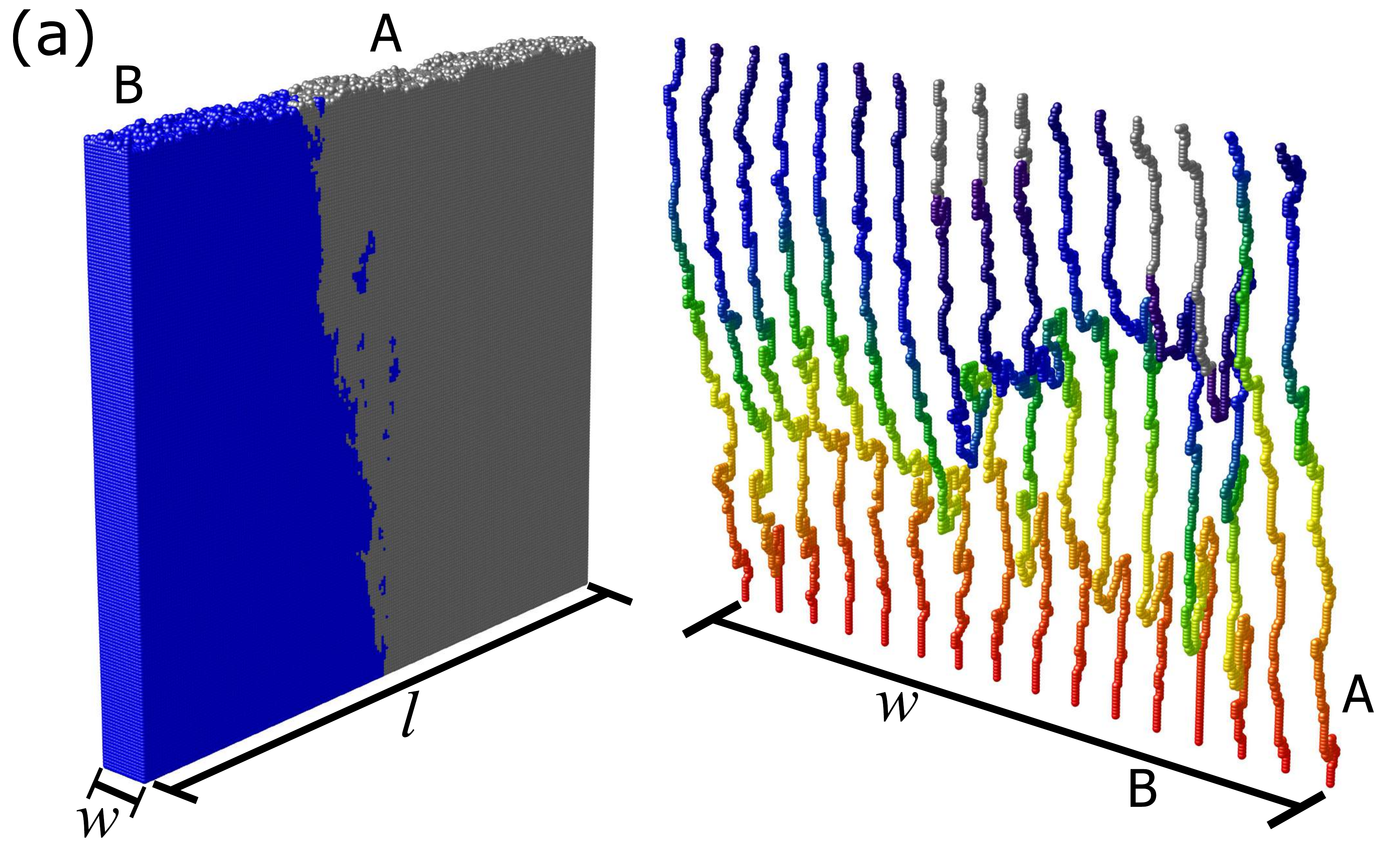}  
		\includegraphics[width=0.25\linewidth]{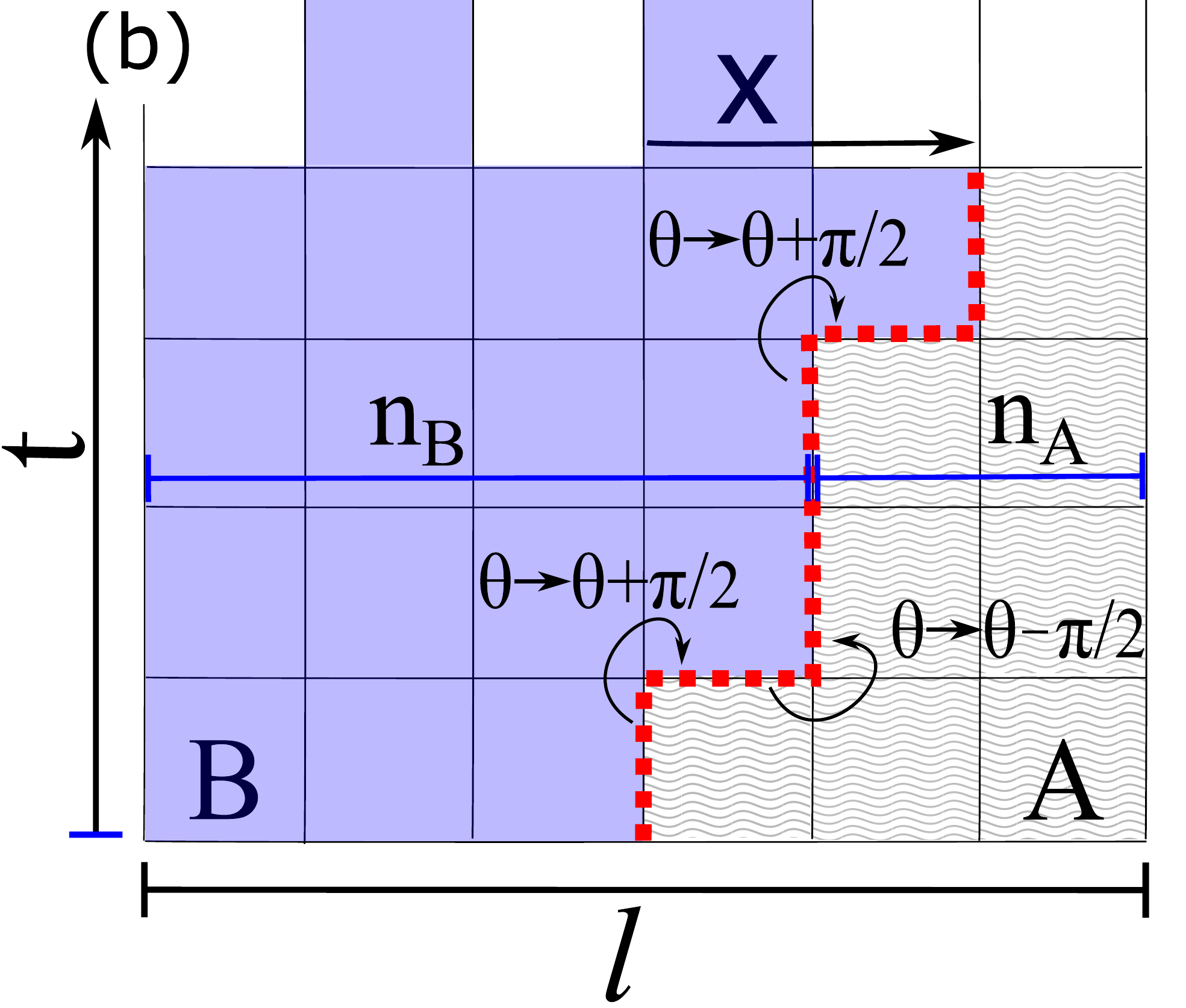}  
		\includegraphics[width=0.32\linewidth]{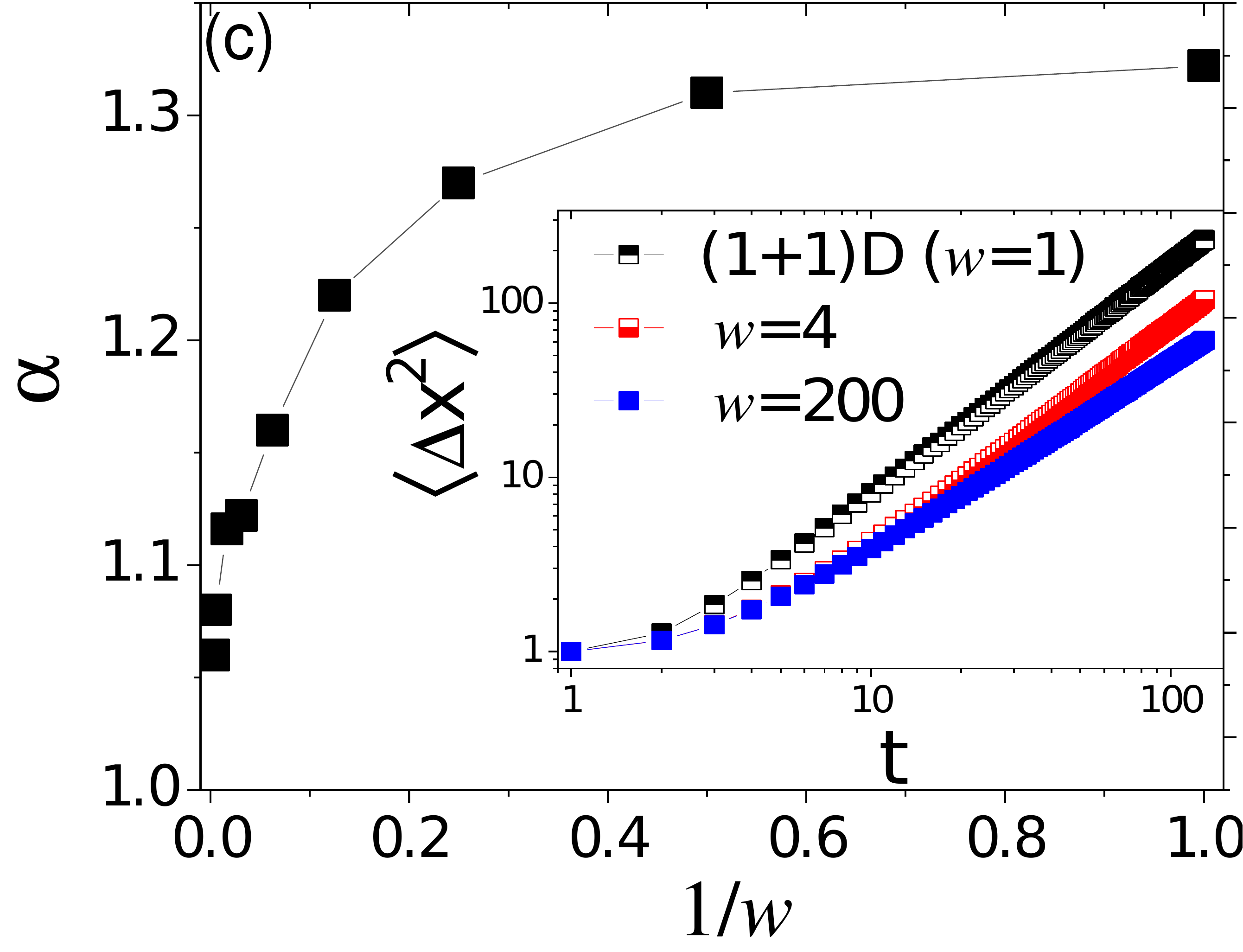}
		\includegraphics[width=0.32\linewidth]{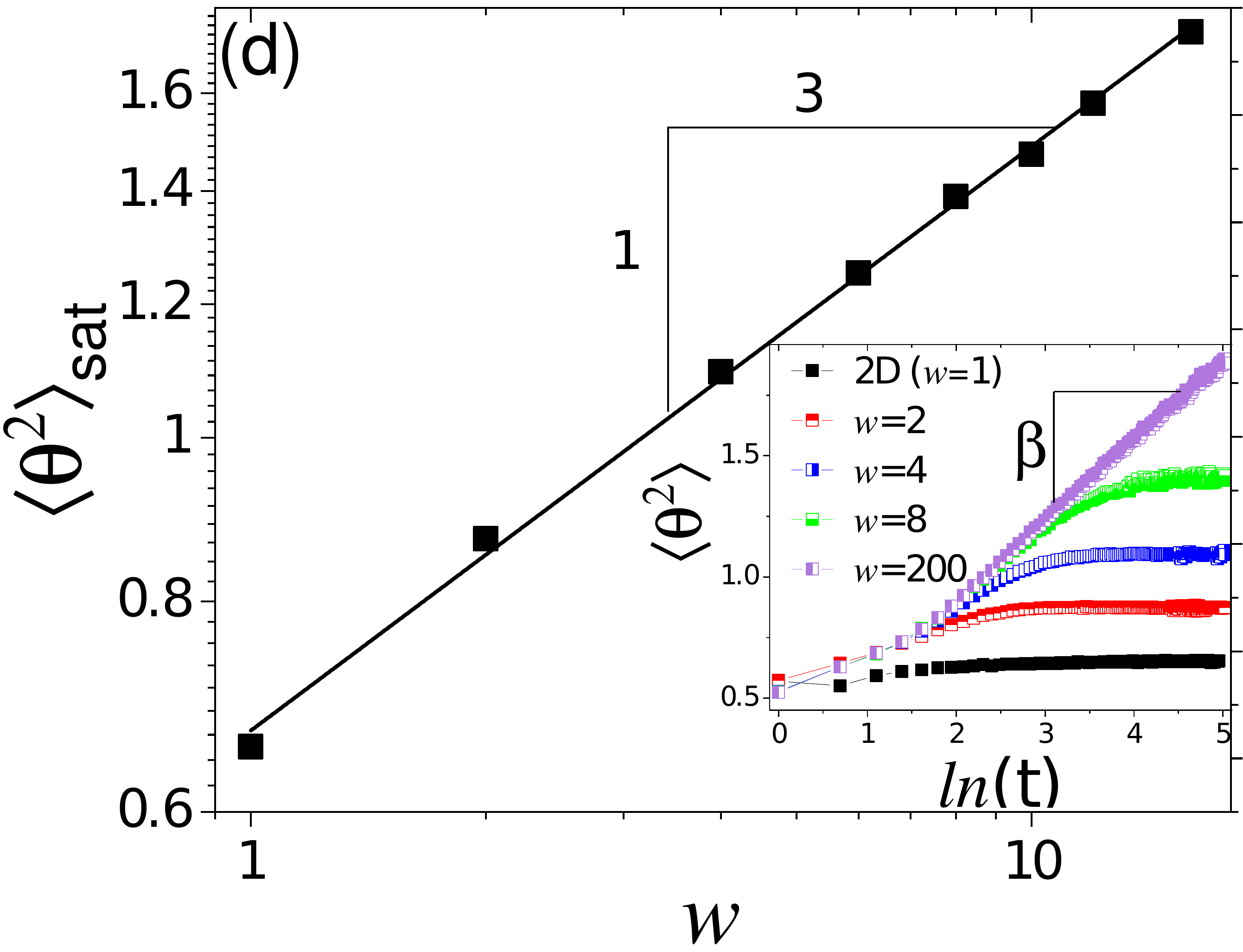}
		\includegraphics[width=0.32\linewidth]{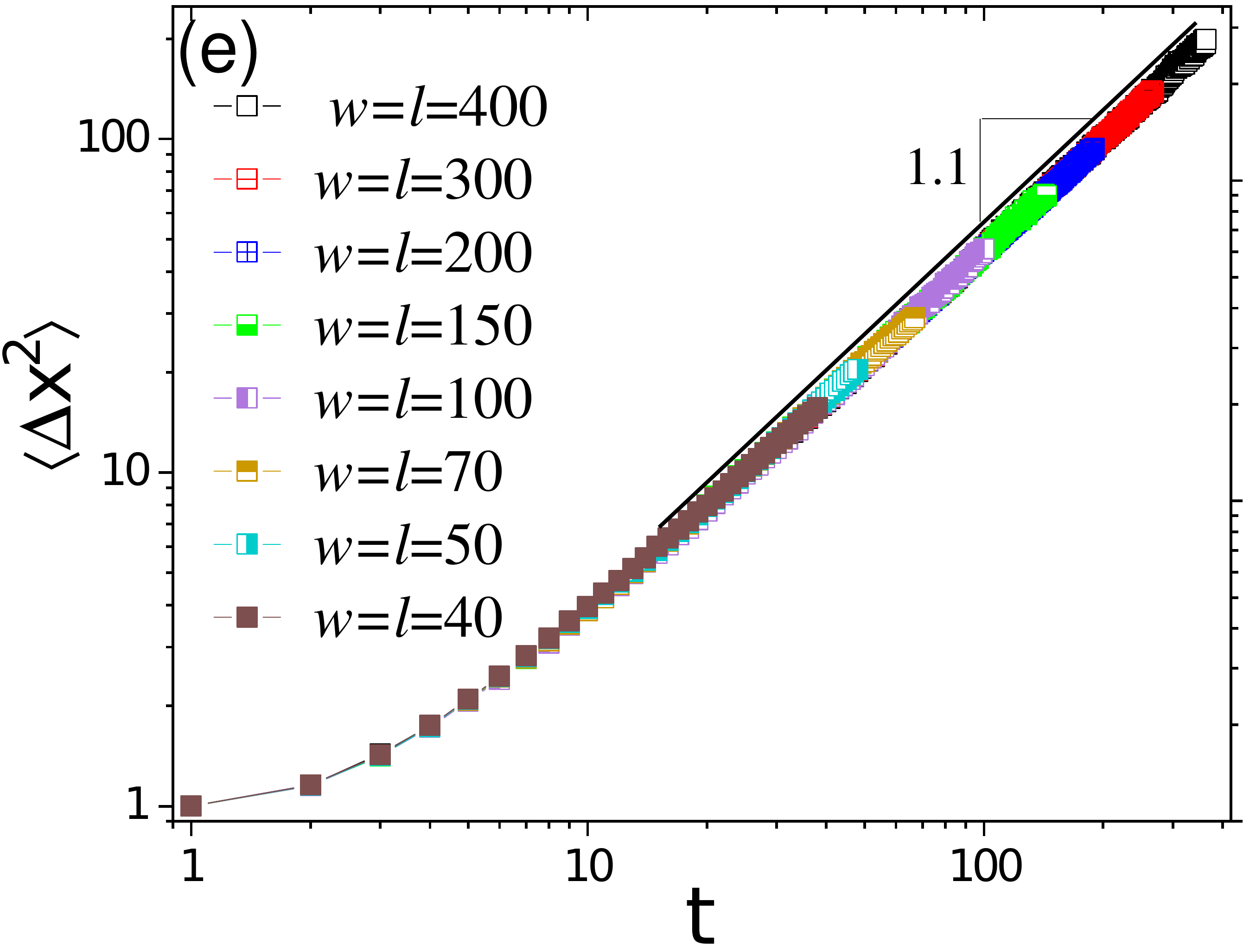}  
		\caption{(a) (2+1)D setting by considering a $w \times l \times l$  environment with $l =200$ and $w=16$ as the environment in which populations grow. We apply temporal evolution until 95 percent of the environment become occupied by cells. A plain consisting of one line at each $yz$ plain, separates two populations and we study the geometry of these lines. (b) For each line, we analyze displacement, $X$, and winding angle, $\theta$. (c) The statistics of displacement of these lines shows $\langle \Delta X^2\rangle \sim t^ \alpha$ (inset). For (1+1)D ($w=1$) we have $\alpha={1.31\pm 0.01}$ and extrapolation suggests to have $\alpha(1/w=0) =1.02$. (d) For statistics of winding angle versus $t$  we have $\langle\theta^2\rangle=a+\beta ln(t)$ with $\beta=0.32\pm 0.01$ until $\langle\theta^2\rangle$ saturates to $\langle\theta^2\rangle_{\text{sat}}$ (inset).  $\langle\theta^2\rangle_{\text{sat}}$ increases versus $w$ as $w ^{0.33\pm 0.02}$.  (e) To check if such a variation in $w$ has dominant effect on domain wall fluctuations or it is just a crossover effect, we compared $ \langle \Delta X^2\rangle $ for environments with different sizes. This analysis shows that the behavior is irrelevant of system size.}
		\label{FIG1}
	\end{figure}

	We apply the temporal evolution in the model, which lets both populations evolve and expand in $z$ (i.e., temporal)-direction. Our simulations are performed until $95$ percent of the environment is occupied by the cells. Such a setting gives rise to the appearance of a single domain wall between the two population kinds whose statistical properties are the main subject of the present study (see FIG. \ref{FIG1}(a)). We dissect this domain wall to constituting lines (one line at each $yz$ plane) and study these lines. Additionally, to have a quantitative understanding of evolutionary processes, we analyze Heterozygosity ($H$), which is defined as $H(t)=n_A (t)\times n_B (t)$ and measures the coexistence of two sub-populations.
	
	\subsection{Domain Walls Geometry.} 
	For each line in a domain wall (total number of $w$ lines in each plain), we analyze two aspects of its geometry, statistics of the displacement $X$ in the $x$-direction and the winding angle $\theta$ (see FIG. \ref{FIG1} (b)). $X$ and its fluctuations provides a quantitative understanding of sub-populations variations along invasion front and can be used to quantify corresponding underlying genetic drift \cite{hallatschek2007genetic, hallatschek2010life}. Winding angle and its statistics shows the local turnings in domain walls and can provide an analytical solution for extinction times  \cite{cardy2005sle, gruzberg2006stochastic}. It is, however, worth mentioning that the initial motivation for computation of the winding angle statistics was to examine if the conformal invariance property holds for the biofilms embedded in 2D. This property is well-known in the context of  2D critical phenomena which provides a very useful and powerful tool i.e., the conformal field theory,  to make exact predictions for the statistical behavior of the 2D random curves which are the domain walls separating different phases of the model.   We use the tie-breaking rule \cite{saberi2009thermal} to define the underlying square lattice's domain walls uniquely. We start the domain walls analysis with (1+1)D ($w=1$). Domain walls displacements in $x$ axis, $X$, exhibit scaling behavior versus time as: $\langle X^{2}\rangle\propto t^{\alpha}$ with $\alpha \sim 1.31 \pm0.01$. Such fluctuations are superdiffusive-like \footnote{This analogy relies on anomalous diffusion studies which suggests fluctuations with $\alpha<1$, $\alpha=1$ and $\alpha>1$ are sub-diffusive, diffusive and super-diffusive. However, for example, $\alpha=1$ should follow other properties of diffusive behavior, thereby being called diffusive. Our analysis revealed that extinction times follow inherently different statistics and, respectively, $\alpha=1$ is diffusive-like.}. As FIG. \ref{FIG1} (c) shows, $\alpha$ decreases as we increase $w$ and extrapolating existing data shows $\alpha(1/w=0) =1.02$. We then analyze winding angle statistics \cite{cardy2005sle, gruzberg2006stochastic}. $\langle \theta^2 \rangle$ increases by $t$ until it approaches $\langle\theta^2\rangle _{\text{sat}}$. For large values of $w$, we have $\langle\theta^2\rangle=a+\beta ln(t)$ with $\beta=0.32\pm0.01$. For saturation limit of winding angle we have  $\langle \theta^2 \rangle _{\text{sat}} \propto w^{0.33\pm 0.02}$.
	
	 Such a difference in domain walls statistics between 2D and 3D environments is not limited to this problem. Across different topics, similar behavior has been observed. A simple example is the Ising model, in which actual 2D systems and 2D snapshots of 3D systems do not exhibit exactly the same dynamics  \cite{dashti2019two}. Additionally, there is a possibility that what we have seen in FIGs \ref{FIG1} (c) and \ref{FIG1} (d) is a system size effect. For such a case,  by varying the system size, a trend should appear in which the larger system sizes have different behavior, similar to what one may expect. However, our analysis (FIG. \ref{FIG1} (e)) shows that the observed behavior is size-independent and moving towards 3D environments genuinely affects domain wall fluctuations.    
	
	\subsection{Extinction Time Analysis.} Despite observing the superdiffusive-like behavior in simulations and experiments, fluctuations of sub-populations considered to be diffusive and, respectively, their extinction times were described by a simple random walker statistics  \cite{hallatschek2010life}. Since we have $\alpha(1/w=0) =1.02$, one may naively assume that for larger values of $w$, extinction probabilities approach to that of a simple random walk and previous assumptions work for (2+1)D settings. We analyzed extinction times for two identical populations ($R_A=R_B$) to clarify this point. When the domain wall touches one side of the defined environment, one of the populations goes extinct, a typical first passage time problem. For (1+1)D case, the dynamics of domain walls has been suggested to follow simple random walk \cite{hallatschek2010life}. Based on this suggestion, for an environment with the size of $l$ extinction probability versus time fits well to a Log-Normal distribution as $P(t)=\frac{A}{\sqrt{2\pi}t\sigma} e^{\frac{[ln(t/\bar{t})]^2} {2\sigma^2}}$ with $\sigma=0.85\pm0.01$ where $\bar{t}$ is the averaged extinction time and we have $\bar{t}\sim l^2$ (see SI).  
	
\begin{figure} 
		\centering   
		\includegraphics[width=0.33\linewidth]{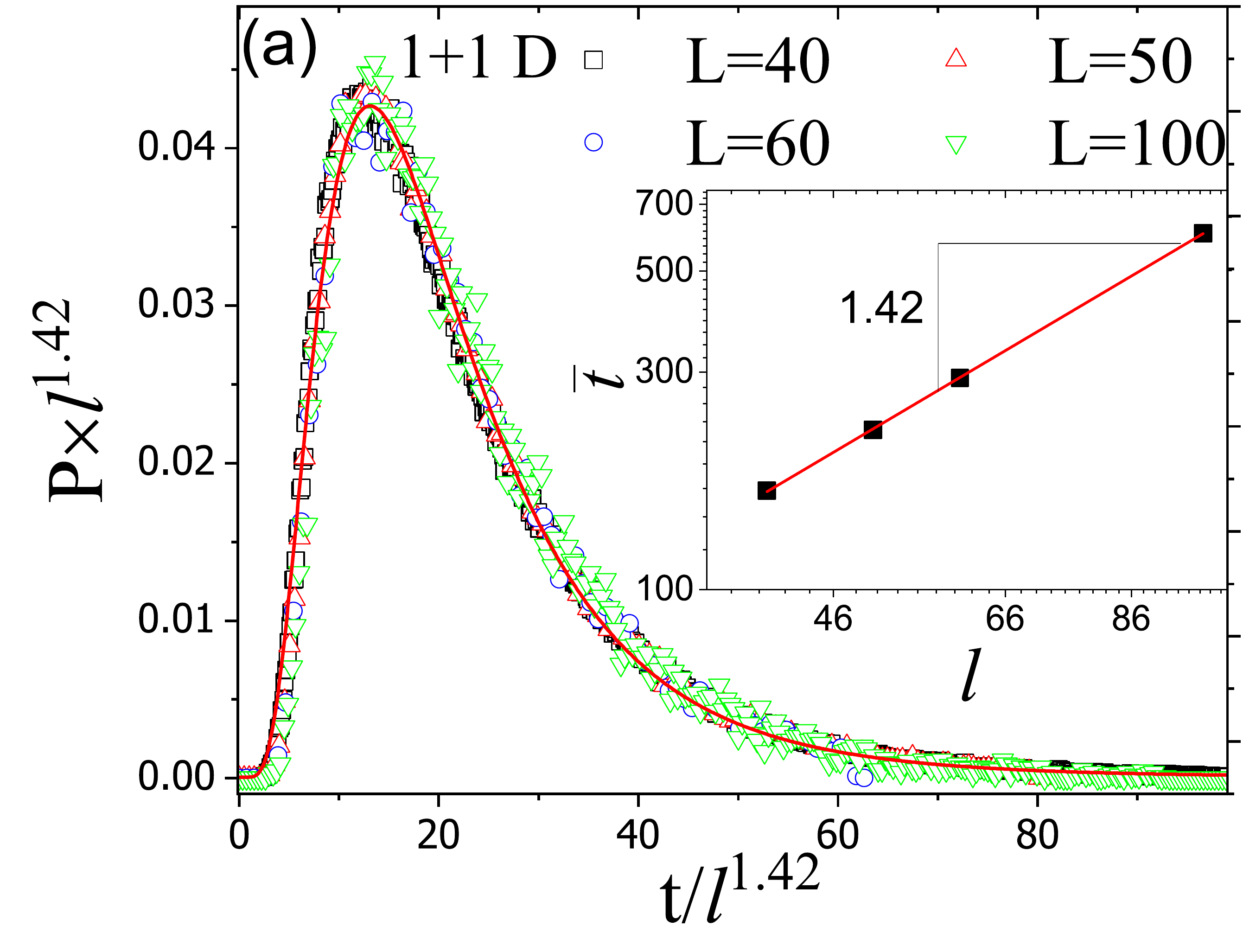} 
		\includegraphics[width=0.33\linewidth]{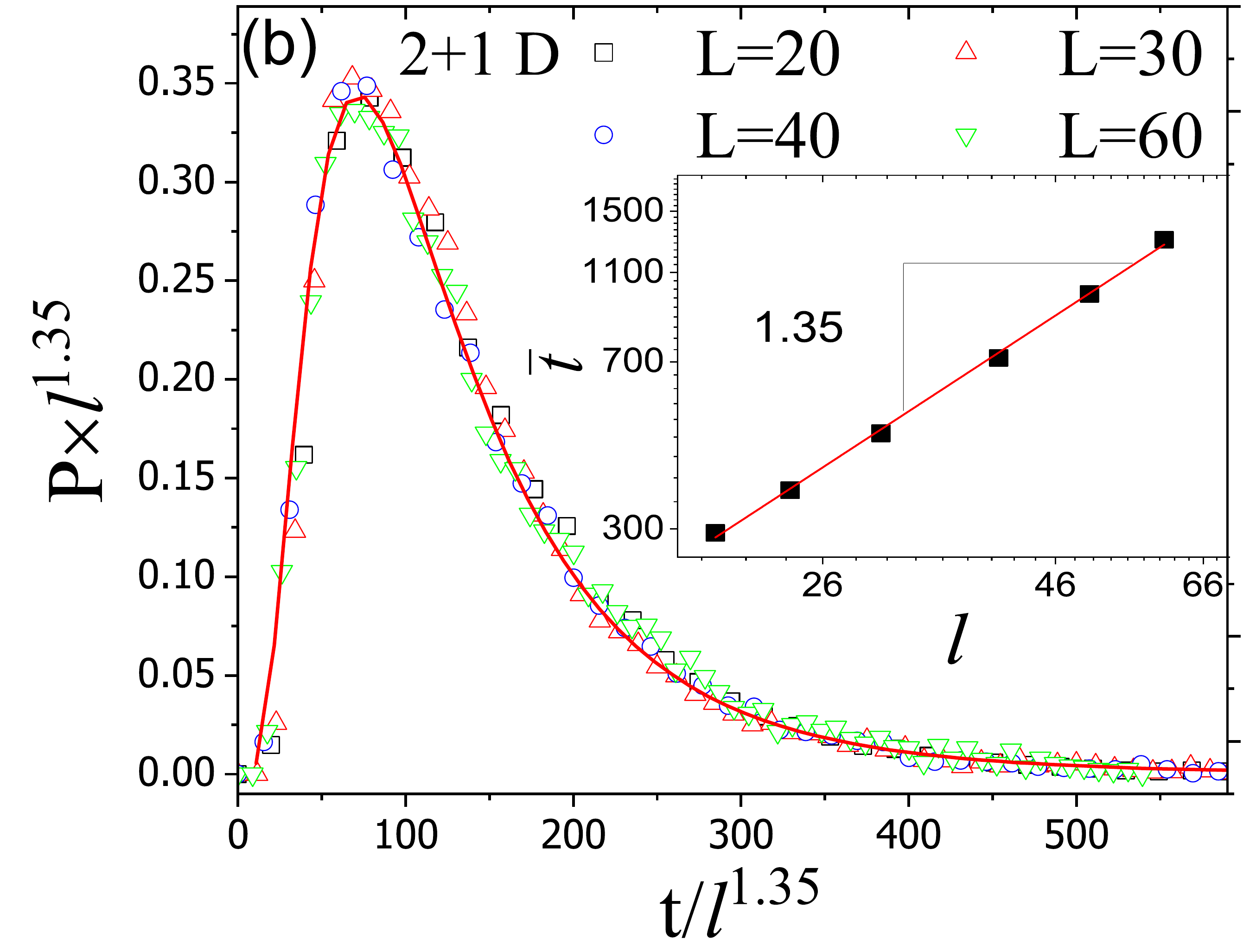}
		\caption{(a) Normalized extinction probability versus normalized $t$ for different system sizes ($l$) in a (1+1)D environment ($w=1$). All sizes show Log-Normal distribution with $\sigma=0.57 \pm 0.01$. Inset: The averaged extinction time grows versus system size as $\bar{t} \sim l^{1.42\pm 0.01}$. These dynamics is significantly different from that of Brownian motion which has been used to analyze extinction during range expansion so far \cite{hallatschek2010life}. (b) The normalized  averaged extinction time versus $t$ for $w=30$ and different values of $l$. Extinction times still have a Log-Normal distribution with $\sigma= 0.54 \pm 0.03$. Inset: The averaged extinction time grows versus system size as $\bar{t} \sim l^{1.35\pm 0.02}$.}
		\label{FIG2}
	\end{figure}

	We start with $w=1$ and run simulations until one of the populations goes extinct and analyzes extinction times statistics. Our results for (1+1)D reveal that while extinction times still follow Log-Normal distribution (see FIG. \ref{FIG2} (a)), the width for these distributions is different and we have: $\sigma=0.57\pm 0.01$. Additionally, the averaged extinction time has a different dynamics versus system size as: $\bar{t} \sim l^{1.42 \pm 0.01}$ (see FIG. \ref{FIG2} (a) inset). These differences suggest that spatially explicit populations may have entirely different dynamics. 
	
	As the next step we studied environments with $w>1$ in which we run simulations until one of the populations goes extinct (all lines in domain wall touch the same side of the environment) and then analyzed statistics of extinction times. For different values of $w$, extinction time distribution remains Log-Normal and width of the distribution remains fixed as $\sigma = 0.54\pm 0.03$ (see FIG. \ref{FIG2} (b) for $w=30$). Interestingly, the averaged extinction time, $\bar{t}$, increases as $\bar{t} \sim l^{1.35\pm 0.02}$, independent of $w$ (see FIG. \ref{FIG2} (b) inset for $w=30$). Thus, while domain walls have a diffusive-like fluctuations with $\alpha \sim 1$, extinction times follow inherently different dynamics (compare $\bar{t} \sim l^{1.35\pm 0.02}$ for $w=30$ with $\bar{t} \sim l^{2}$ for simple random walk). In light of this finding, we checked if fluctuations of domain walls in flat-front models are identical to a simple random walker. Using our model for flat-front growth \cite{azimzade2019short}, we found that while extinction times follow a Log-Normal distribution with $\bar{t} \sim l^{2}$, distributions width is not the same as simple random walker and we have $\sigma=0.72 \pm 0.02$. Respectively, even domain walls in flat-front models exhibit diffusive-like fluctuations (and not diffusive). Finally, the averaged extinction time  increases versus $w$ as $\bar{t} \sim w^{0.50}$.
	
	\subsection{Wedged Initial Condition.} 
\begin{figure}  [!h]
		\centering
		\includegraphics[width=0.32\linewidth]{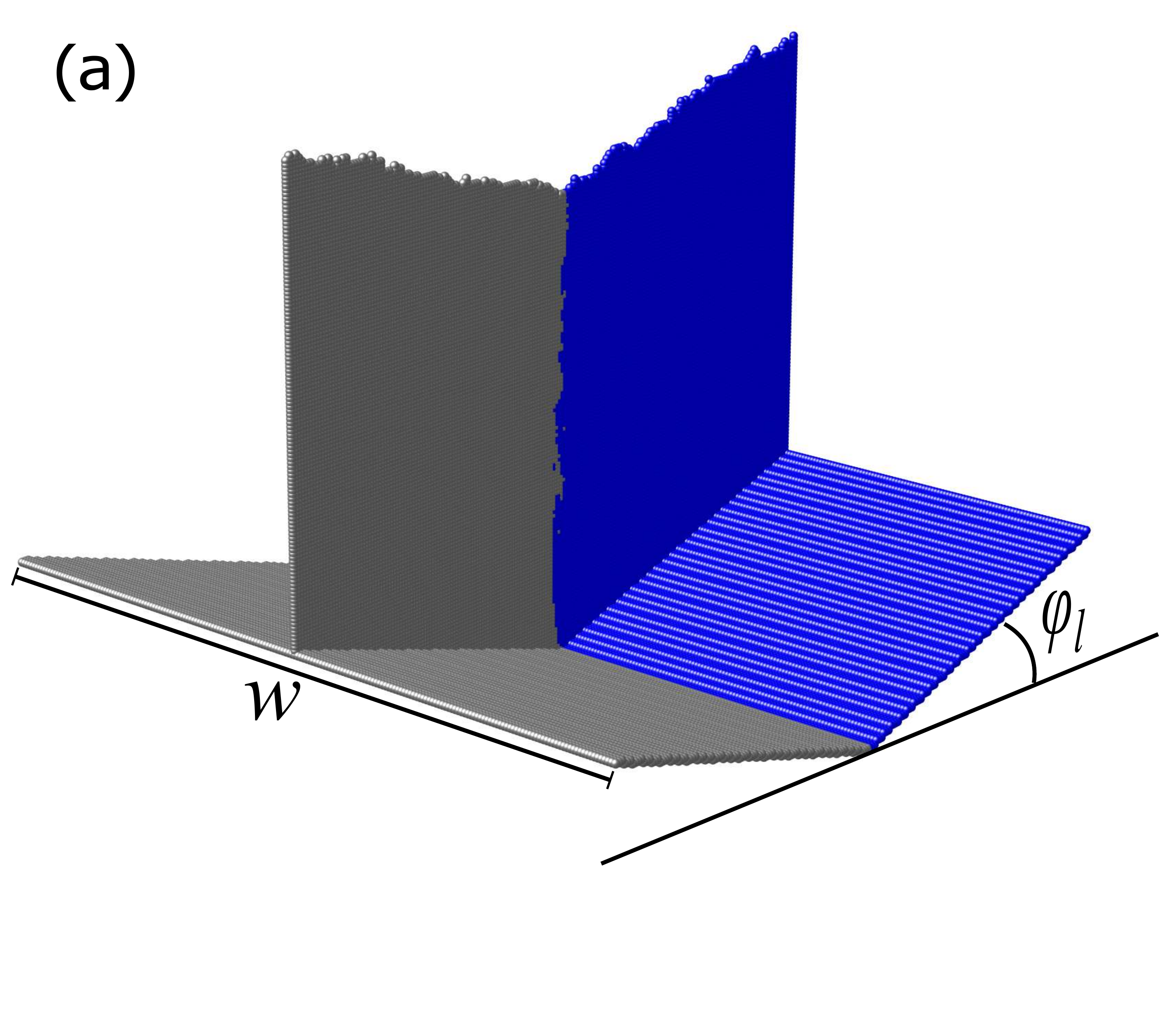}  
		\includegraphics[width=0.32\linewidth]{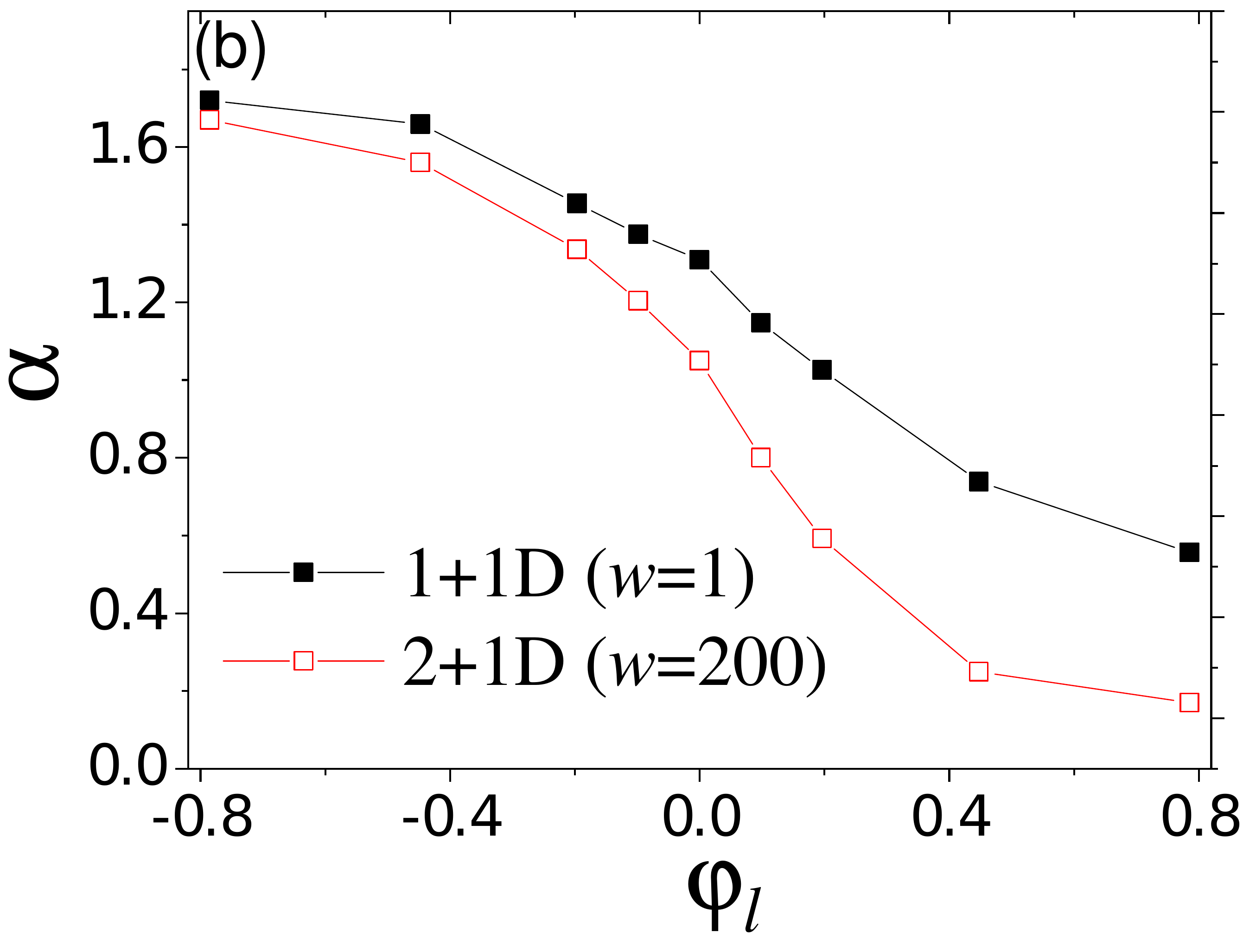} 
		\includegraphics[width=0.32\linewidth]{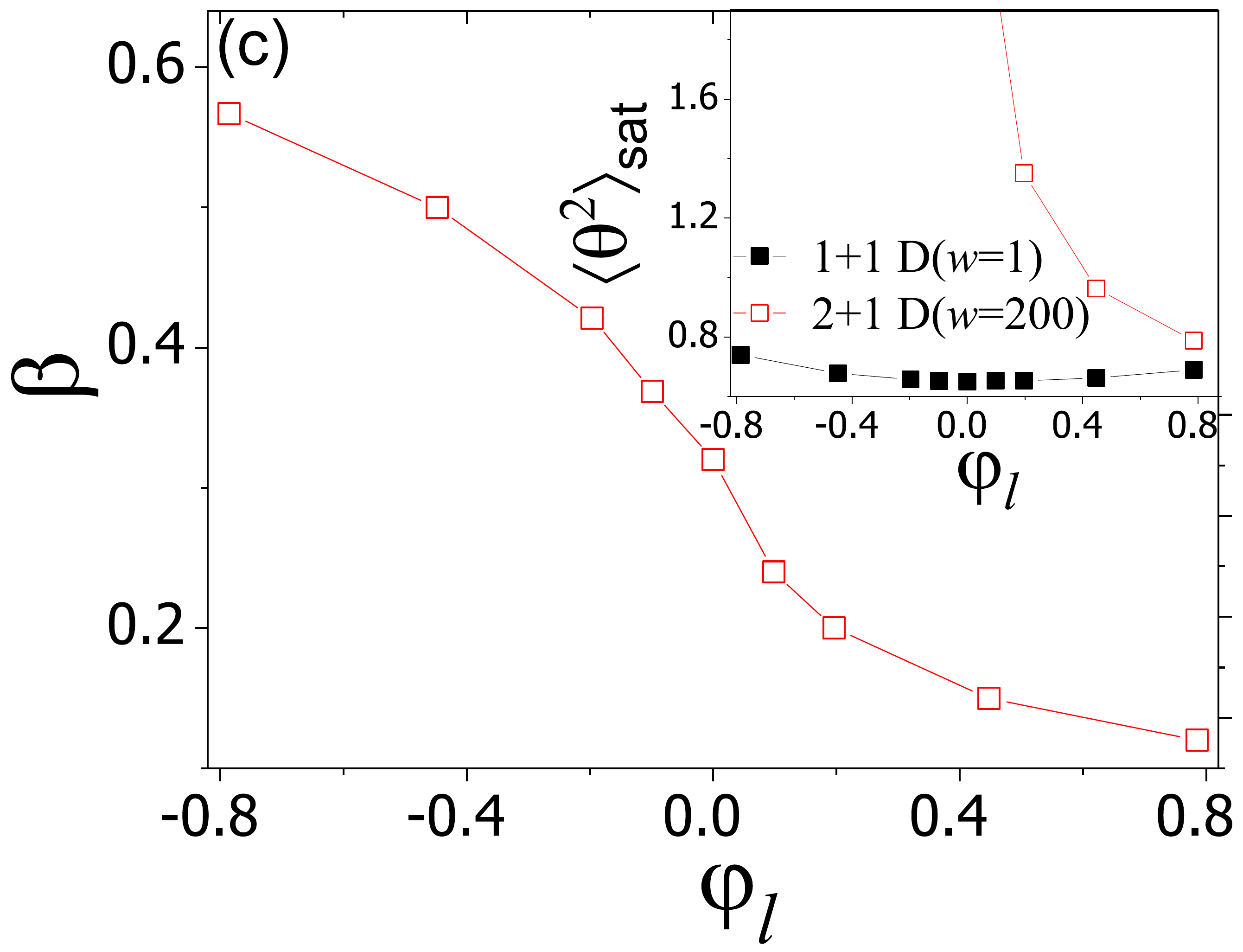}   
		\caption{(a) (2+1)D wedge setting as shown with angles of $\phi_l$. (b) $\alpha$ versus $\phi_l$ for (1+1)D and (2+1)D. As we increase $\phi_l$, $\alpha$ monotonically decreases. (c) $\beta$ versus $\phi_l$. $\beta$ decreases versus $\phi_l$. Inset: $\langle\theta^2\rangle_{\text{sat}}$ versus $\phi_l$ for (1+1)D and (2+1)D. These results together reveal that initial condition can have longstanding effect on the geometry of domain lines.}
		\label{FIG3}
	\end{figure}

Biofilms normally form on surfaces that are not flat. It has not been understood if the surface (substrate) geometry can have a relevant effect on the evolutionary dynamics of growing biofilms. To implement a fabricated geometry on the surface, we assume a different initial condition. Similar to the flat initial condition depicted in FIG. \ref{FIG1} (a), each population initially lives on a half-strip, and two half-strips touch each other at $(x=0,z=0)$ line with the angle of $2\phi_l$ as shown in FIG. \ref{FIG3} (a). It should be noted that $\phi_l=0$ represents the flat initial condition setting studied before and for the case of (1+1)D   half-strips become half-lines.
	
	Again, we calculate two geometrical features of lines: displacement and winding angle statistics. Starting with analysis of $\alpha$ for 1+1D case, our results show that increasing $\phi_l$ from $-\pi/4$ to $\pi/4$ monotonically decreases $\alpha$ from $1.62$ to $0.42$ (see FIG. \ref{FIG3} (b)) which is in agreement with previous models \cite{derrida1991interface, chu2019evolution}. As the next step we analyze $\alpha$ for $w>1$. Results for $w=200$ reveals that the monotonic behavior is independent of $w$. We also analyze the winding angle statistics for different values of $w$. $\langle\theta^2\rangle$ monotonically decreases as we increase $\phi_l$. Additionally, $\beta$ monotonically decreases versus $\phi_l$ as shown in FIG. \ref{FIG3} (c). $\langle\theta^2\rangle _{\text{sat}} $ also shows a monotonic decrease versus $\phi_l$ for $w>1$ (see FIG. \ref{FIG3} (b) inset).  
	
	These results suggest that the initial condition affects the geometry of domain walls. As indicated before, the dynamics of sub-populations depend on the geometry of these domain walls. Naturally, the question arises about whether or not the initial condition affects the growing population's evolutionary dynamics? 
	
	\subsection{Wedged Initial Condition on Evolutionary Dynamics.} 
\begin{figure} [!h]
		\centering   
		\includegraphics[width=0.32\linewidth]{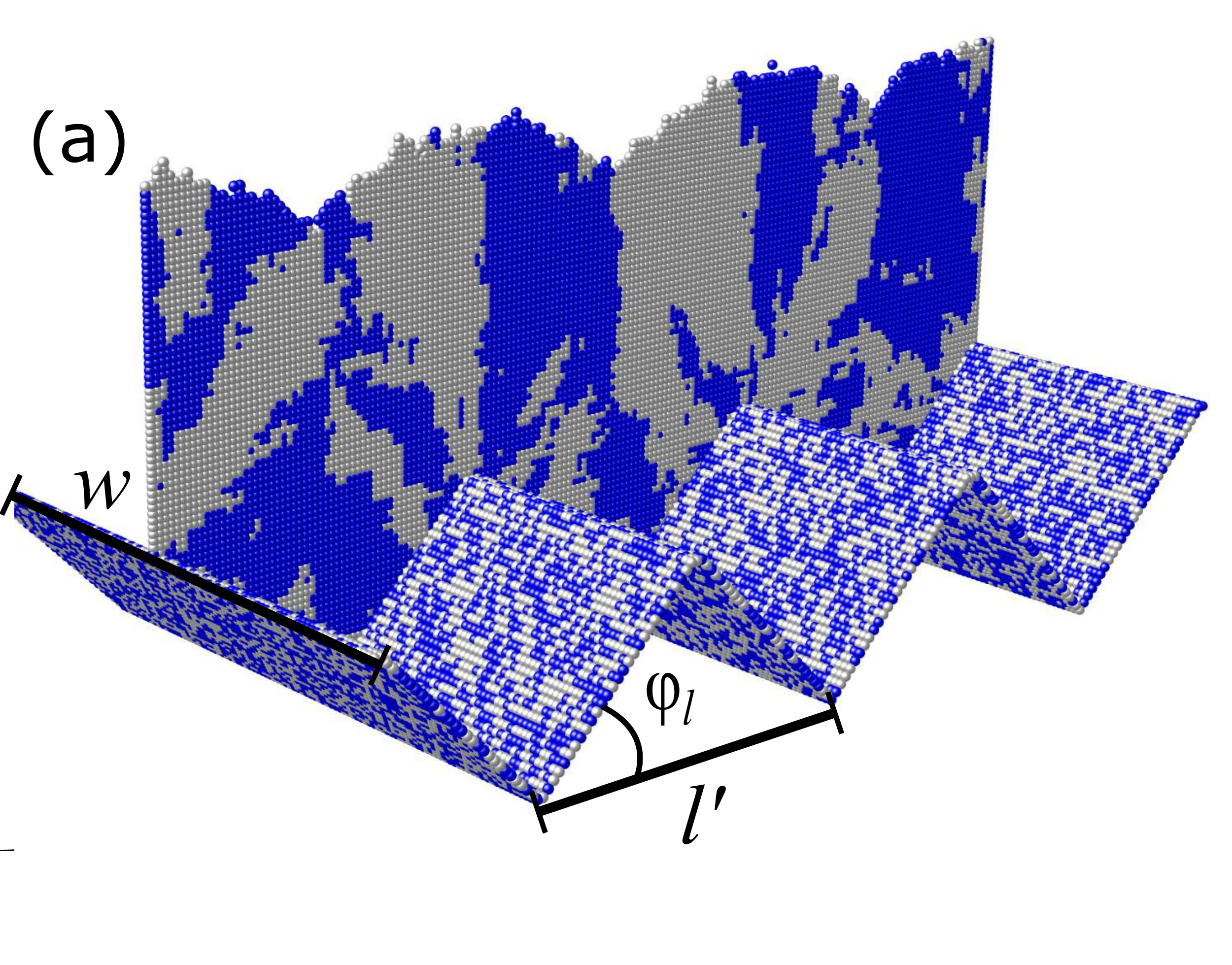} 
		\includegraphics[width=0.32\linewidth]{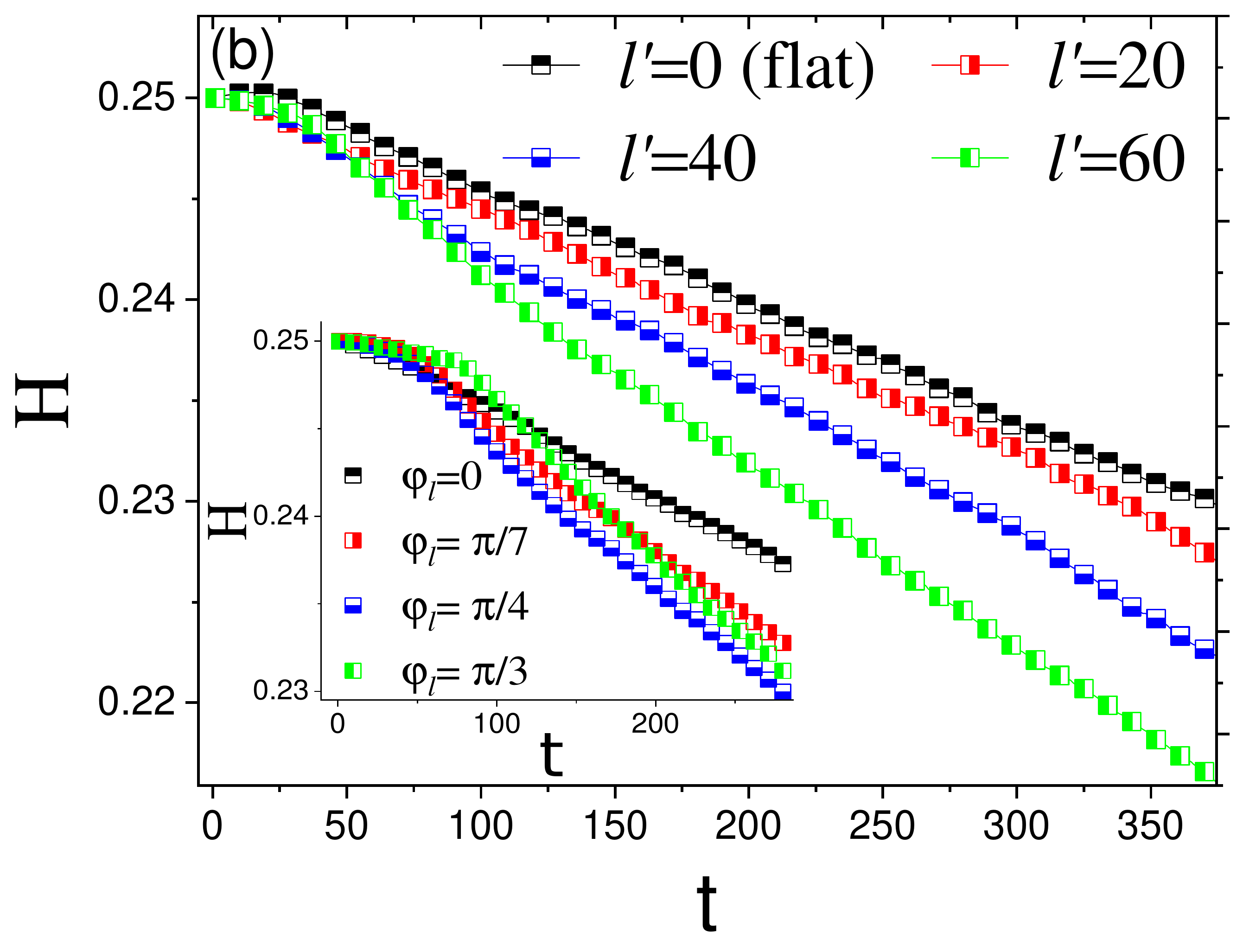}  
		\includegraphics[width=0.32\linewidth]{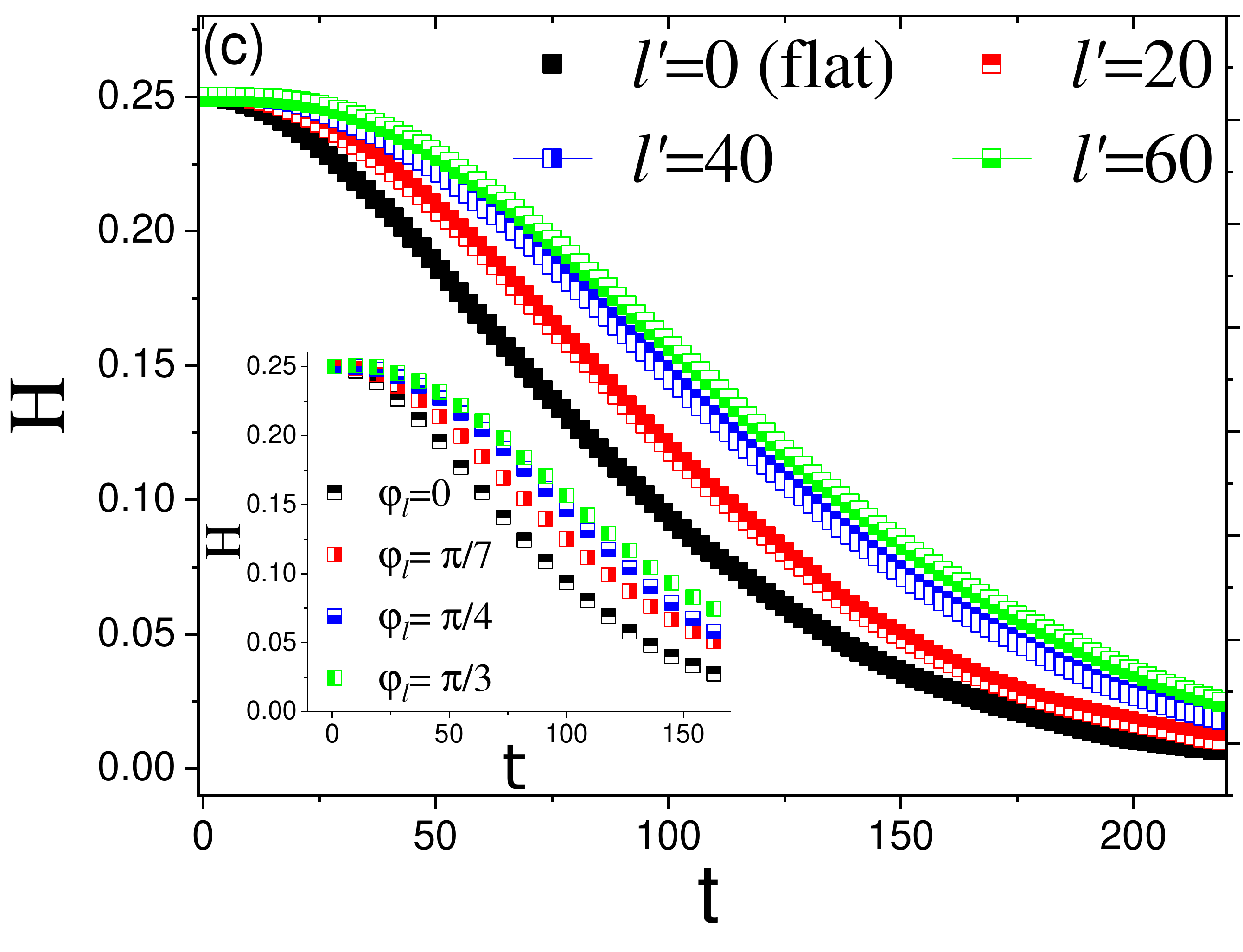}     
		\caption{(a) A $yz$ cut of a growth on periodic wedged initial condition with angle of $\phi_l$ and periodic length of$l'$. Instead of initially separated populations, we consider two well mixed populations on a periodic wedge initial condition. (b) Heterozygosity, defined as $H(t)=n_A\times n_B$, versus $t$ for different values of $l'$ with $\phi_l= \pi/4$ and $w=100$ (The qualitative behavior is independent of $w$). Inset: $H(t)$ versus $t$ for different values of $\phi_l$ and $l'=40$. Thus, initial condition can interfere fluctuations of sub-populations and one can accelerate genetic drift through engineering the geometry of substrate. (c) $H$ versus $t$ where for  $R_A= 1.05 \times R_B$  with $\phi_l= \pi/4$ and  $w=100$ (The qualitative behavior is independent of $w$). Inset: $H(t)$ versus $t$ for different values of $\phi_l$ and $l'=40$. Due to lower fitness advantage, (B) has a lower chance to survive and goes extinct over time. However, as we increase $l'$, (B) gets more chance to survive. Changing initial configuration leads to long-lasting effect on diversity of expanding populations and promotes their co-existence.}
		\label{FIG4}
	\end{figure}

From FIG. \ref{FIG3} we know that one can modulate the geometry of domain walls by manipulating initial conditions. To understand the effect of this fabrication on evolutionary dynamics, we analyze the Heterozygosity of two co-evolving populations (defined as $H(t)=n_A(t) \times n_B(t)$). We assume that both populations are randomly placed on a periodically wedged surface with a periodic length of $l'$ and the angle of $\phi_l$ (see FIG. \ref{FIG4} (a)).
	
	For populations with identical duplication rates, $R_A=R_B$, genetic drift is the only process involved in evolutionary dynamics. As time passes, spatial segregation appears (see FIG. \ref{FIG4} (a)). Analysis of Heterozygosity for different values of $\phi_l$ and $l'$ suggests that the initial condition has a long-lasting effect on the two populations' composition. In the presence of wedged initial conditions, Heterozygosity decreases faster. As such, one of the populations dominates the invasion front faster (see FIG. \ref{FIG4} (b)). This result suggests that biofilms composed of subpopulations with similar fitness advantages that grow on periodically wedged surfaces will have lower diversity.
	
	To further clarify the effect of initial configuration on evolutionary dynamics, we include two non-similar populations as well. To this end, we suppose that one of the populations has a higher duplication rate and set (A) to duplicate faster by only five percent ($R_A=1.05 \times R_B$). As one may expect, (A) dominates the invasion front by time and $H$ declines, respectively. We start with a similar initial condition as described before (shown in FIG. \ref{FIG4} (a)) and run the simulations for different values of $l'$ and $\phi_l$. FIG. \ref{FIG4} (c) shows that increasing $l'$ leads to higher $H$ versus time. In other words, the wedged initial condition allows the weaker population to survive over longer periods. The qualitative behavior for both scenarios is independent of $w$.
	
\begin{figure} [!h]
		\centering   
		\includegraphics[width=0.334\linewidth]{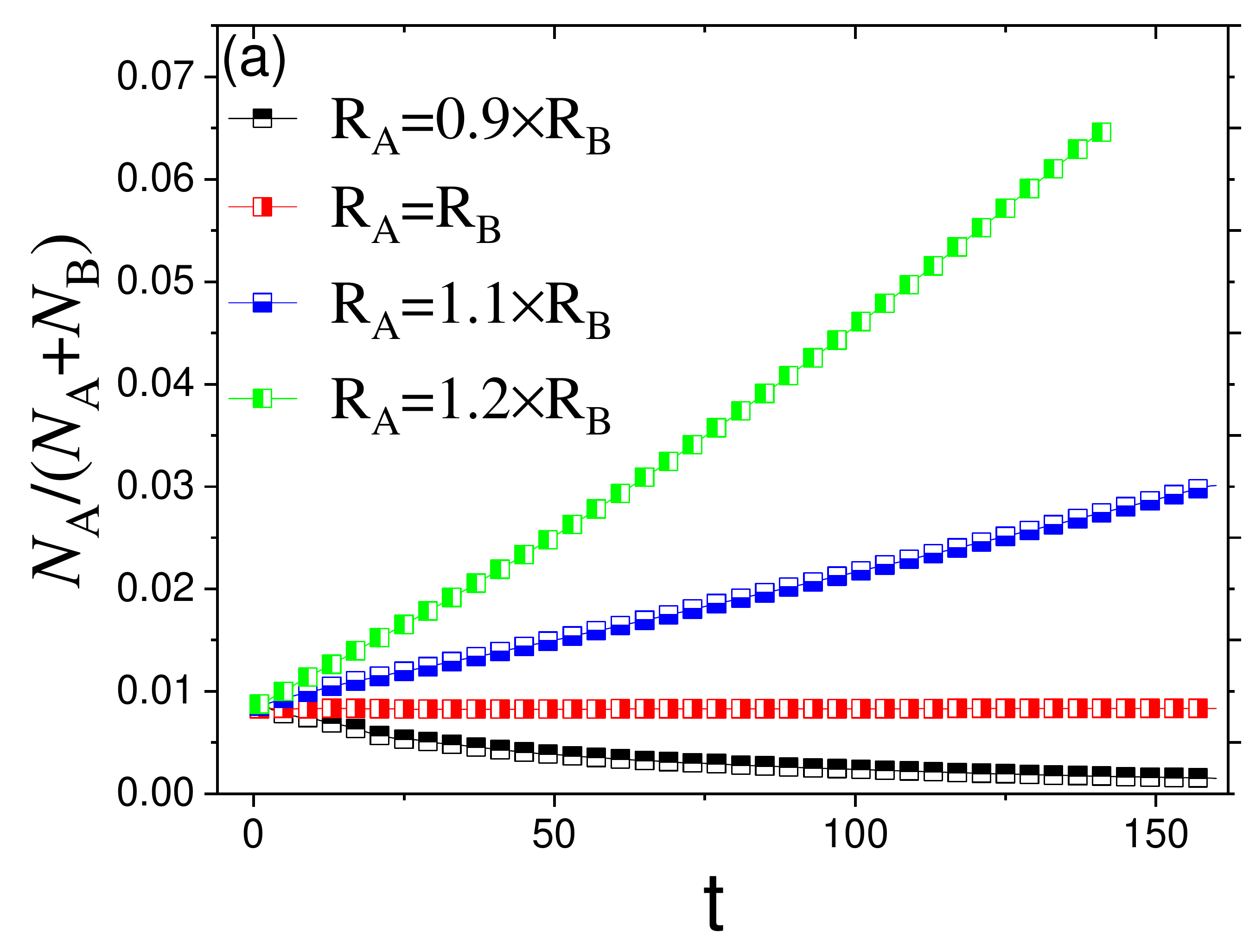} 
		\includegraphics[width=0.334\linewidth]{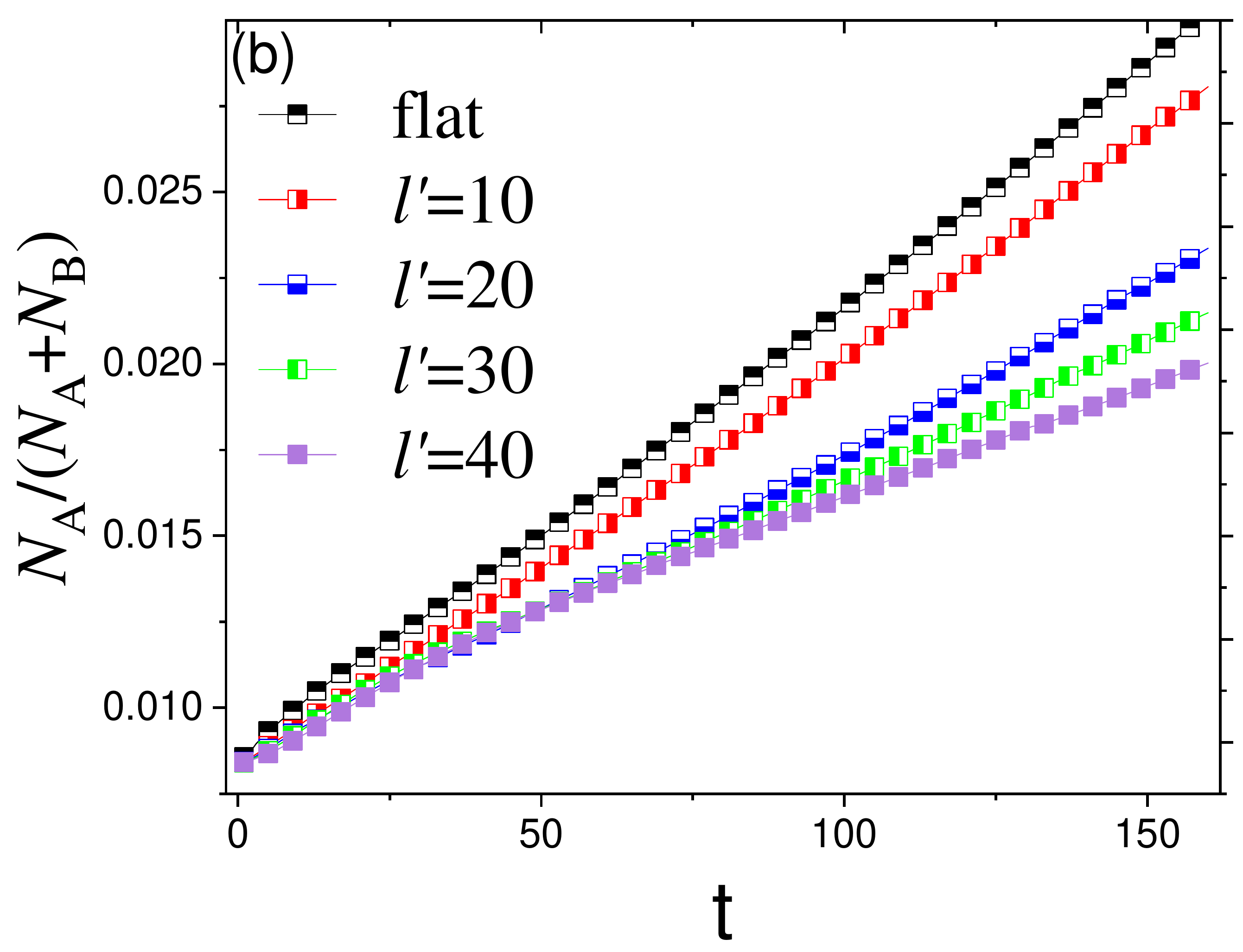}  
		\includegraphics[width=0.334\linewidth]{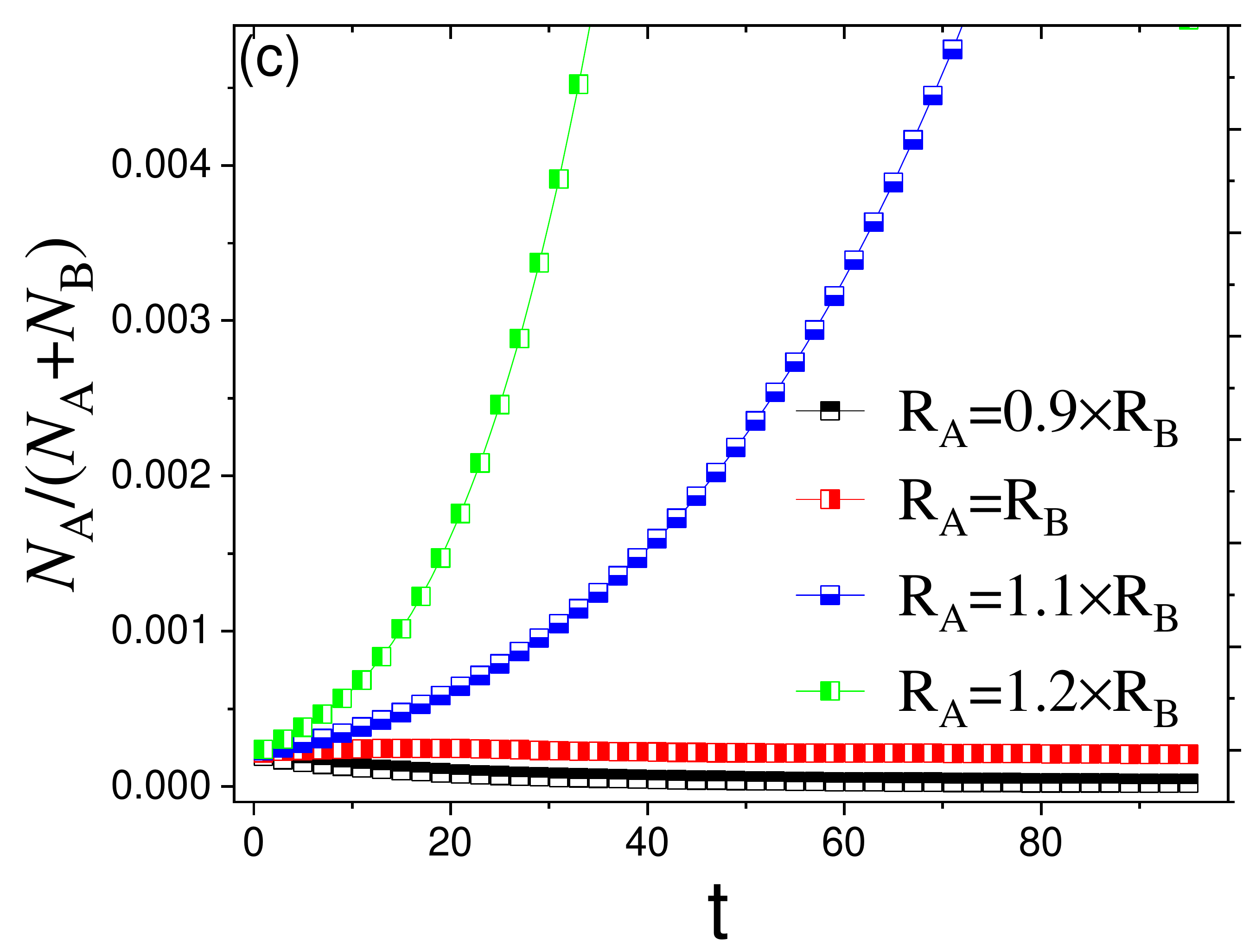}  
		\includegraphics[width=0.334\linewidth]{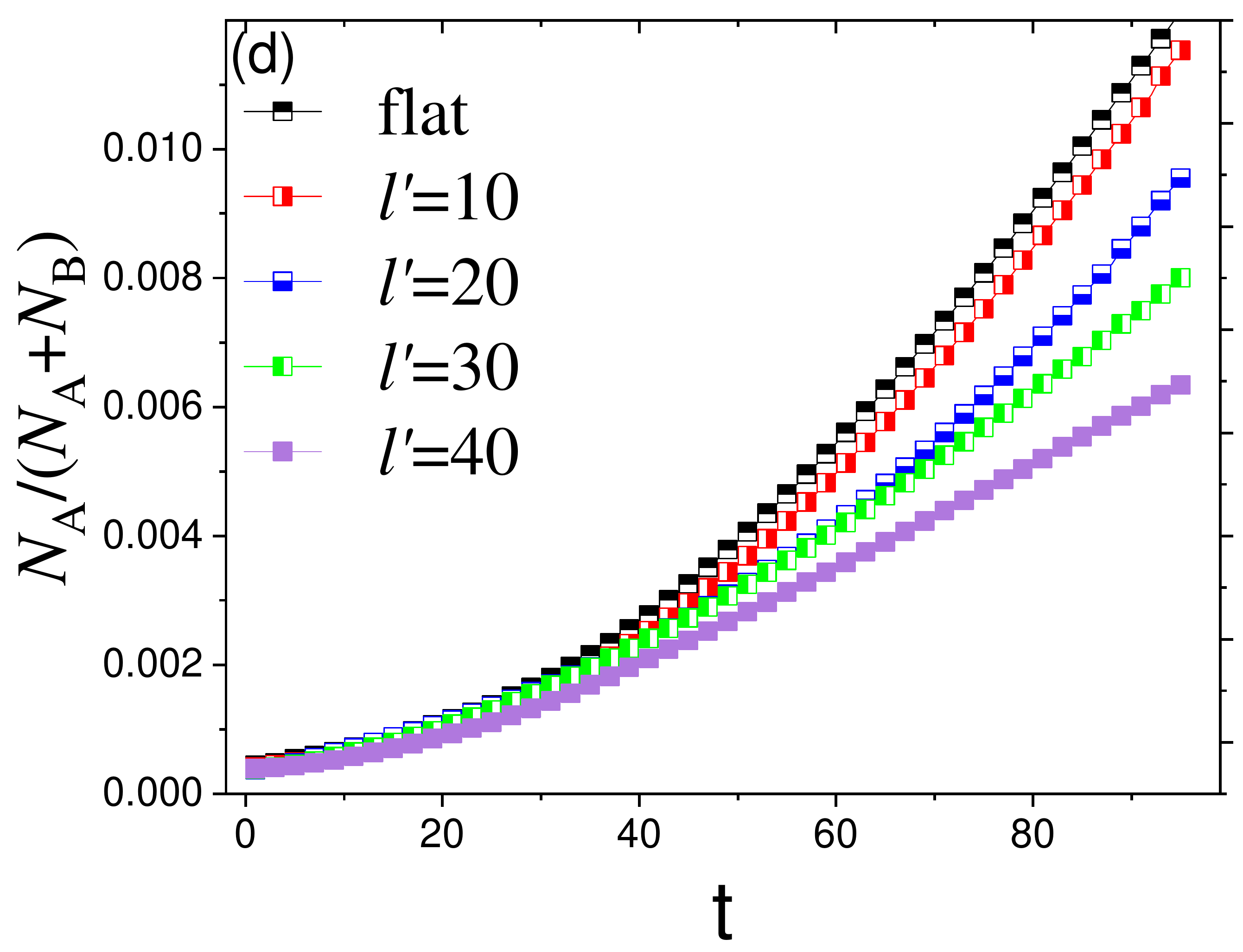}    
		\caption{ (a) A single member of (A) emerges in a 2D environment which already is occupied by (B). Due to mutations, (A) can have a different duplication rate as $R_A=R_B(1+\Delta R)$ where $ \Delta R<0$,  $ \Delta R=0$ and  $ \Delta R>0$ represents deleterious, passenger and driver mutations, respectively. Such difference in duplication rate affects the probability of fixation for (A). As we increase $\Delta R$, the fixation probability increases.  (b) For beneficial mutants, living on a 2D periodically wedged surface with $\phi_l=\pi/4$ delays the fixation process, depending on the value of $l'$. This result reveals that the initial geometry of substrate directly regulates the fixation probability of beneficial mutants. (c) and (d) show  a similar process happens in 3D environments as well. }
		\label{FIG5}
	\end{figure}

	 As shown in FIG. \ref{FIG4}, considering a patterned structure for the substrate can interfere with the composition of growing populations. To check how beneficial mutant fixation is affected by such a patterning, we perform additional analysis. We assume that population (B) is already living on a patterned surface, as shown in FIG. \ref{FIG4} (a). A new mutant, namely (A), appears in this environment in the beginning. If the mutant has the same duplication rate ($R_A=R_B$), the chance for fixation of this mutant is equal to each member of (B) and the average fraction of (A) in the environment, $\frac{N_A}{N_A+N_B}$, remains the same over time for both 2D and 3D environments (see FIGs \ref{FIG5} (a) and (c)). If (A) poses a deleterious mutation that had decreased its duplication rate ($R_A=R_B(1+\Delta R)$ with $\Delta R<0$), (A) goes extinct gradually. If (A) posses a driver mutation ($R_A=R_B(1+\Delta R)$ with $\Delta R>0$), its fraction increases over time until it becomes fixed at the expense of (B). In an environment with a flat structure, the magnitude of fitness advantage is the determinant factor. However, on a patterned surface, the fixation also depends on the geometry of the surface. For both 2D and 3D environments, such patterns can delay fixation process (see FIGs \ref{FIG5} (b) and (d)). 
	
	\section{Discussion.} 
	To interfere with the evolutionary dynamics of an evolving population, different strategies have been developed \cite{nourmohammad2019optimal, bittihn2017suppression,  iram2020controlling, gatenby2020integrating}.  For a population composed of subpopulations with similar fitness advantages, a less diverse population is preferred. Such a population has a lower chance of developing resistance \cite{girvan2005bacterial}. On the other hand, when there is a sub-population with higher fitness, slowing down the domination helps other sub-populations survive and have a smaller fitness advantage population. The superior sub-population may even go extinct as a result of random events \cite{gatenby2020integrating}. Since biofilms function depends on the sub-population they contain, there has been a growing demand for regulating biofilms' composition to avoid resistance and/or hinder the dominance of types with higher fitness advantage \cite{mira2015rational, bittihn2017suppression, fitzgerald2019bacterial, santos2019evolutionary, shibasaki2020controlling, sanchez2020directed}.
	
	Here, we first show that spatially explicit populations have inherently different dynamics. They do not follow the  simple random walk, even when domain walls exhibit diffusive-like fluctuations. This observation emphasis that spatial structure changes the way populations interact and evolve \cite{durrett1994importance}. It also paves a way to control population dynamics exploiting their spatial arrangement. 
	
	The initial condition can have long-lasting effects on fluctuations of the invasion front \cite{saberi2019competing, azimzade2019effect}. Since the geometry of invasion front fluctuations and fluctuations of sub-populations are inter-related \cite{hallatschek2007genetic, azimzade2019short}, manipulating invasion front fluctuations may interfere with the evolutionary dynamics of growing sub-populations.  Surface structure affects bacterial attachment, but the possible effect on growing populations' evolutionary dynamics has remained unclear \cite{champigneux2019effect, majhi2020modulating}. We showed that initial geometry could regulate domain wall fluctuations. Inspired by our finding, we studied the effect of the substrate's geometry on growing populations' evolutionary dynamics. Our results revealed that the substrate's wedged structure accelerates the genetic drift, leading to lower diversity among growing populations. Since less diverse communities are less prone to develop drug resistance, such an intervention is desired in biofilms. Additionally, our results reveal that the initial condition favors populations' coexistence for populations with different fitness advantages, leading to slower dominance of types with higher duplication rates. Interestingly, such an intervention is also highly desired.
	
	\section{Conclusion.} 
	Search for parameters that affect population dynamics is of practical interest, particularly due to intervention with evolutionary dynamics. Here we study how the spatial structure of the environment affects evolutionary dynamics. Our results suggest that while ignored so far, the environment's geometry can enforce distinct dynamics on sub-populations fluctuations, affecting evolutionary dynamics in compactly growing populations. Additionally, it can be used to alter the diversity among growing populations when needed.
	
	\section{Acknowledgment.} We would like to thank Mehran Kardar for reading our manuscript and his comments.

\newpage
\section{Supplementary Material}

\section{System Size Effect} 
\begin{figure} [h]
	\centering
	\includegraphics[width=0.38\linewidth]{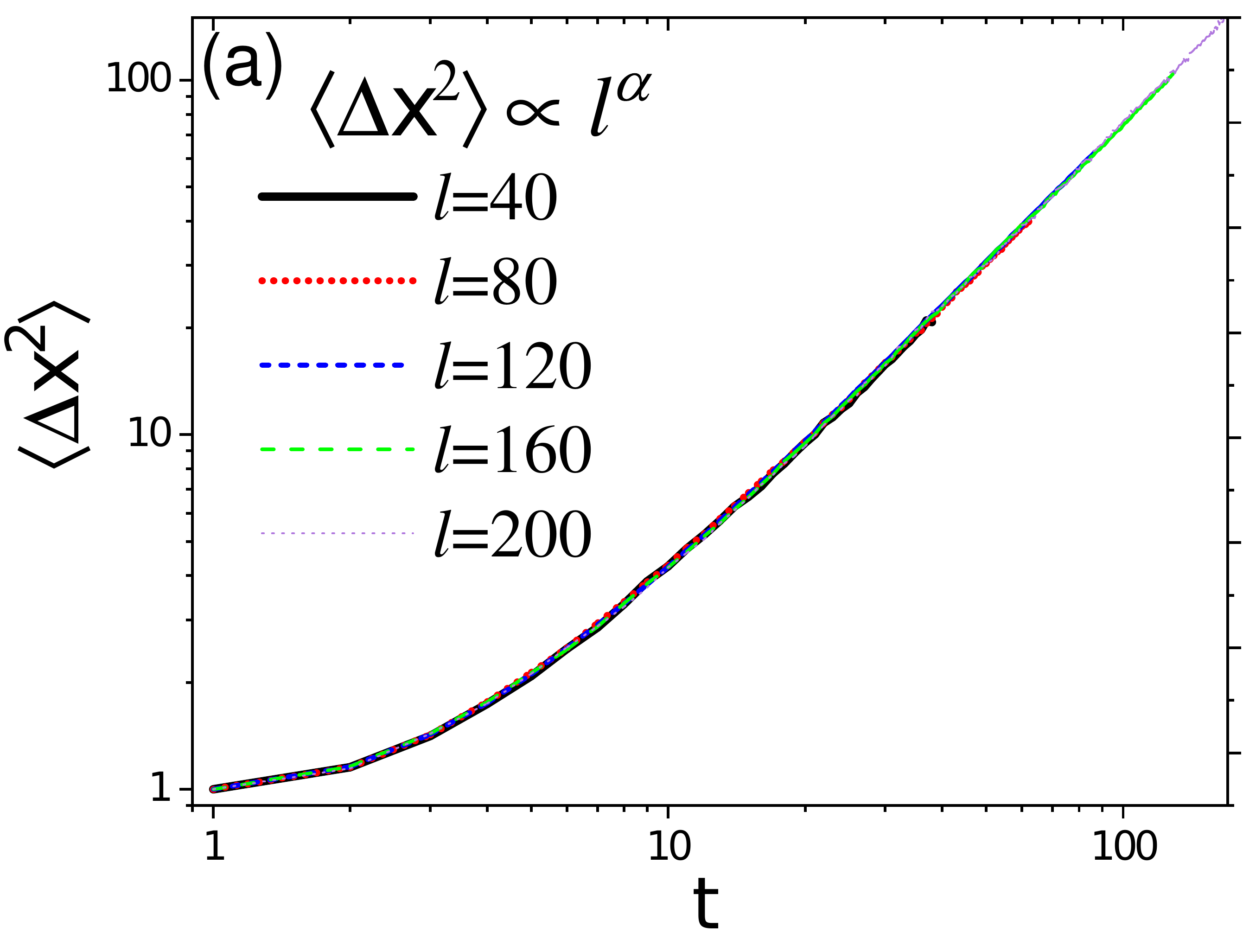}  
	\includegraphics[width=0.38\linewidth]{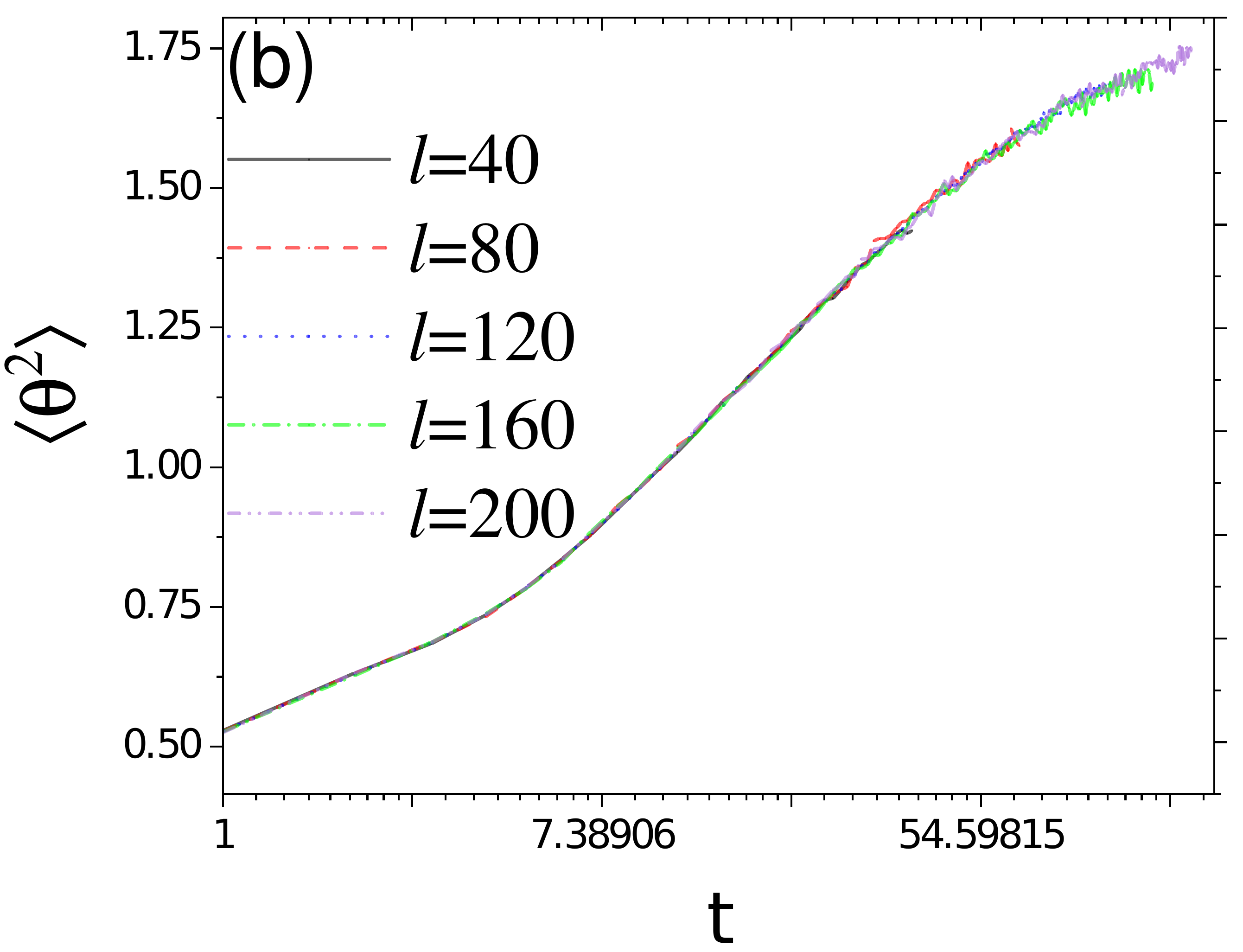}    
	\caption{ system size effect on   $\langle  X^2\rangle$ (a) and  $\langle\theta^2\rangle$ (b)    }
	\label{FIGS1}
\end{figure}

As the first we check how environment size, $l$, affects fluctuations of $x$ and $\theta$. As FIG. \ref{FIGS1} shows, system size seems to be an irrelevant parameter in this regard.  

\section{Extinction Time Analysis}
\begin{figure} [h]
	\centering   
	\includegraphics[width=0.45\linewidth]{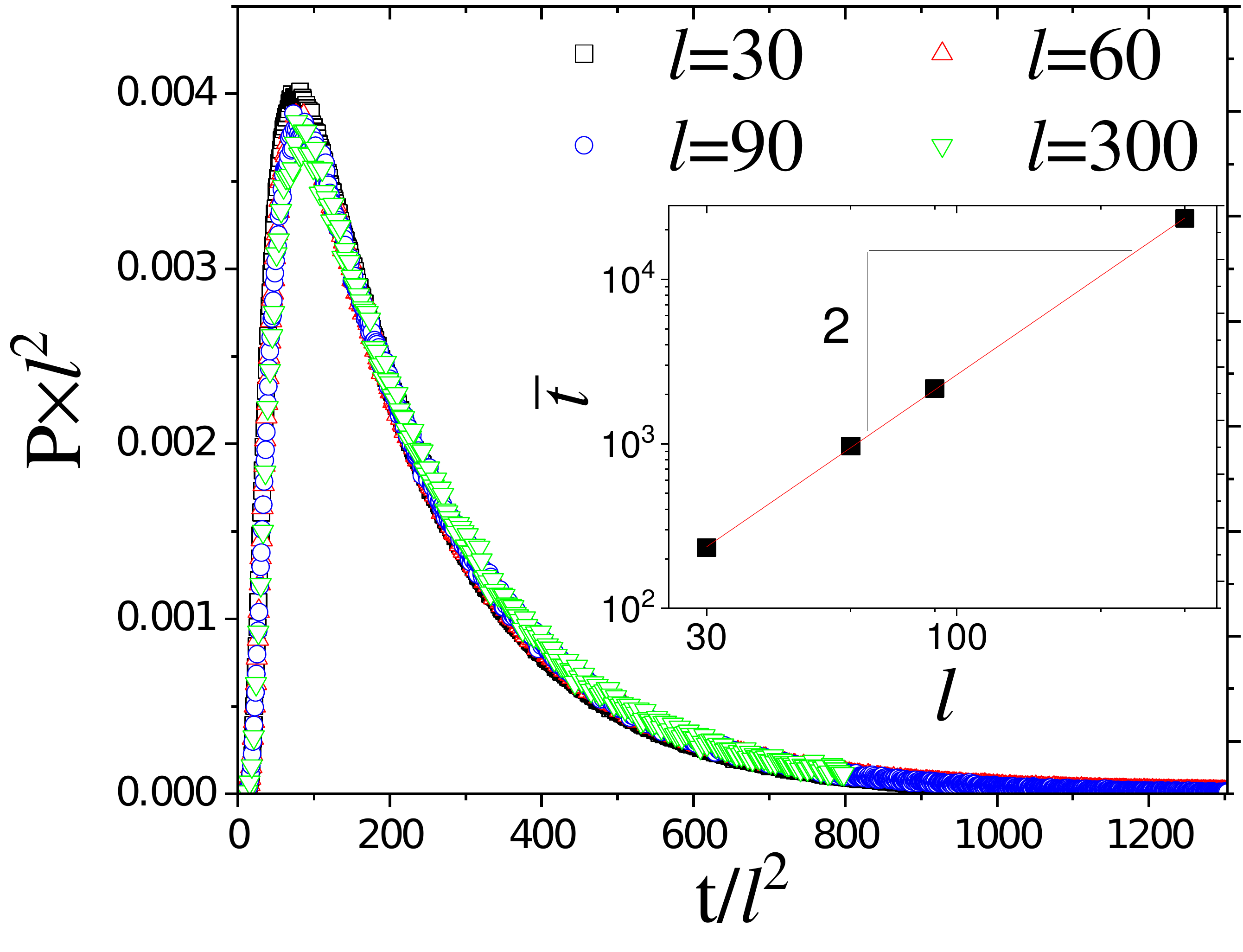}     
	\caption{Extinction probability versus time  of a simple random walker on a lattice of size $l$ for different system sizes. All sizes show Log-Normal distribution with $\sigma= 0.85 \pm 0.01$. Inset: The averaged survival time grows versus system size as $\bar{t} \sim l^{2.00\pm 0.01}$. }
	\label{FIGS4}
\end{figure}

For a simple random walker on a lattice with the size of $l$ the distribution of hitting (extinction) times follows a Log-Normal distribution. Our analysis shows that for such distribution  as $P(t)=\frac{A}{\sqrt{2\pi}t\sigma} e^{\frac{[ln(t/\bar{t})]^2} {2\sigma^2}}$ we have  $\bar{t}\sim l^2$ and $\sigma=0.85$ (see FIG. \ref{FIGS4}).

\section{wedged Initial Condition}
\begin{figure} [h]
	\centering
	\includegraphics[width=0.60\linewidth]{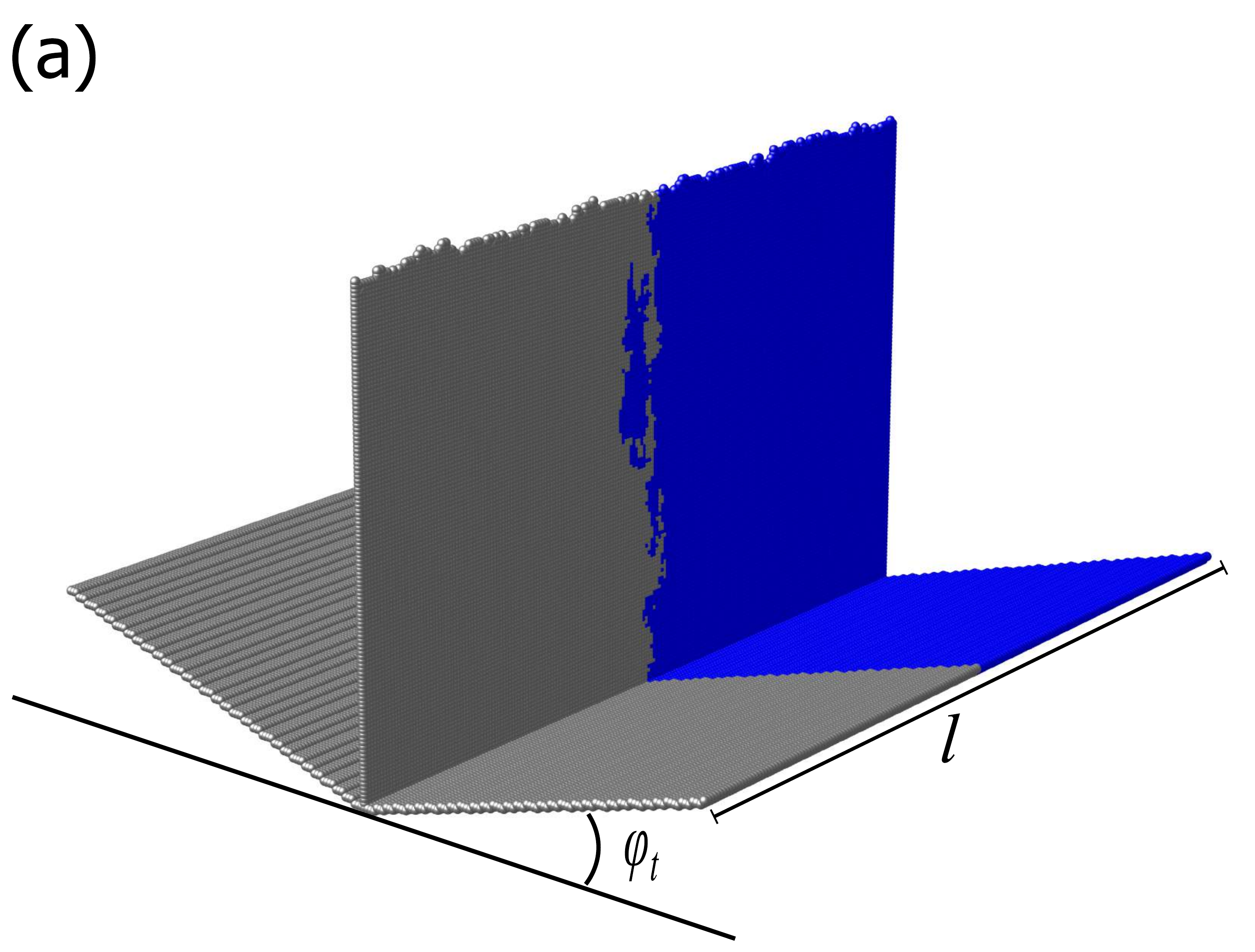}  
	\includegraphics[width=0.40\linewidth]{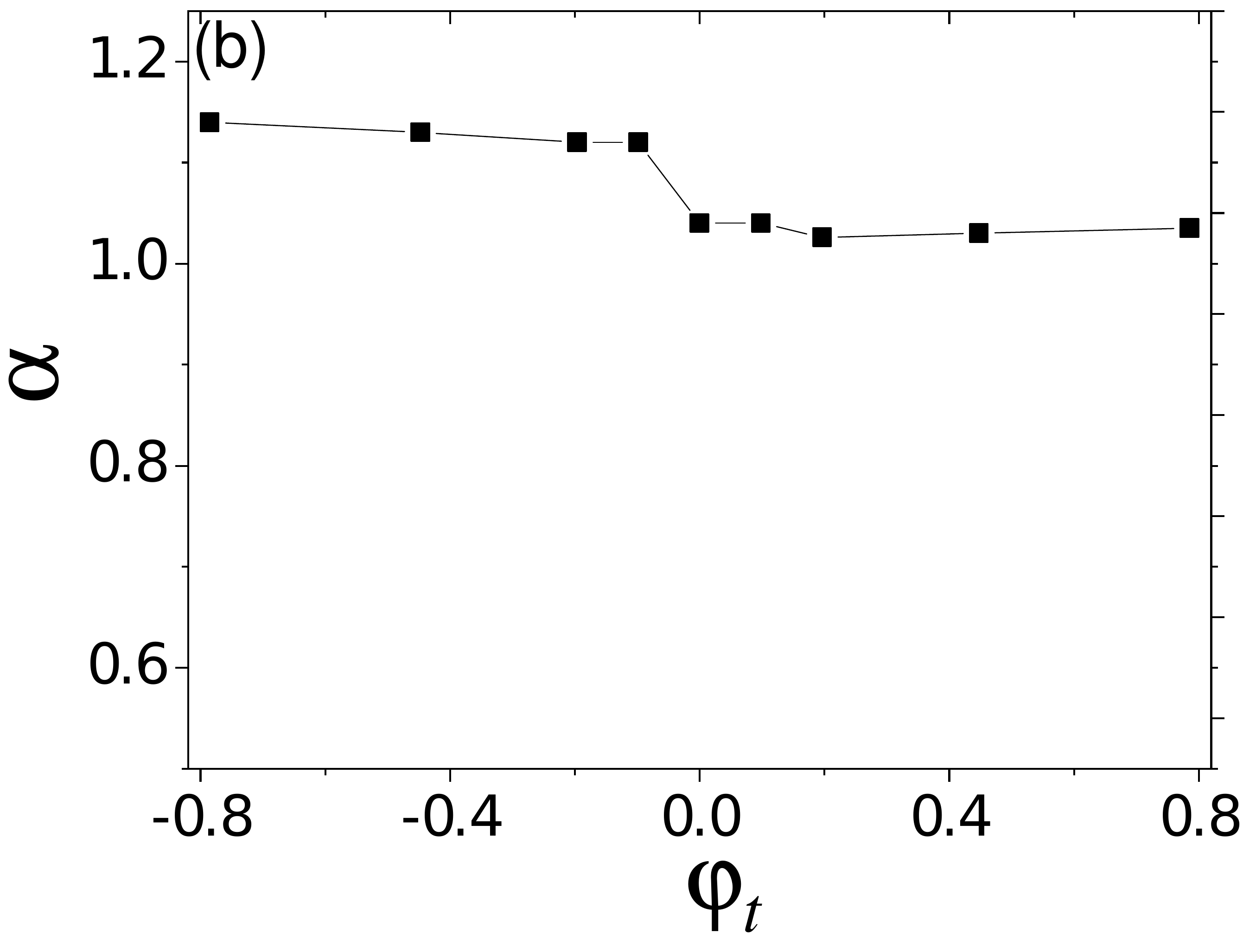} 
	\includegraphics[width=0.40\linewidth]{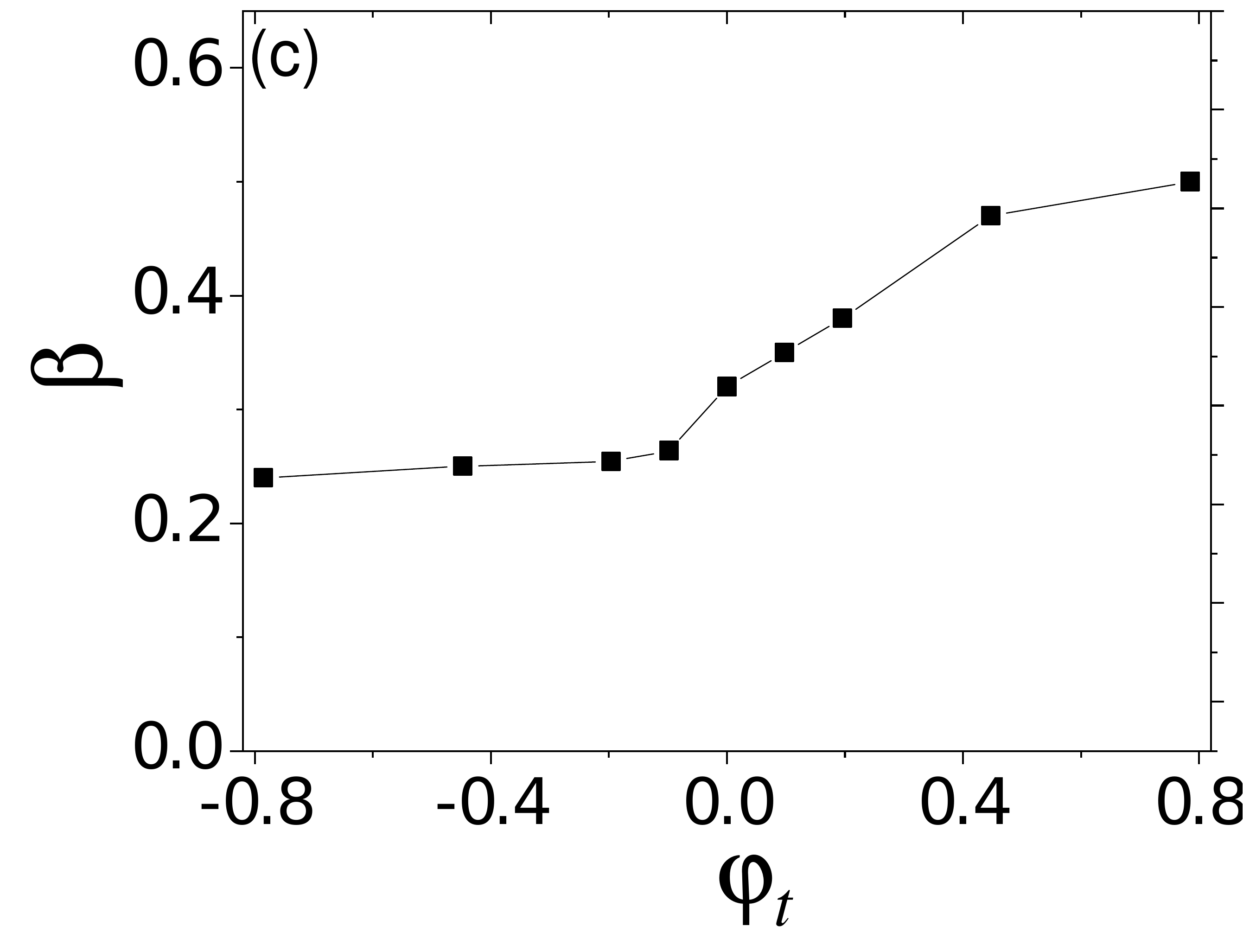}   
	\caption{ (2+1)D wedge setting with angle of $\phi_t$.  (b) $\alpha$ versus $\phi_t$. Increasing $\phi_t$ leads to smaller values of  $\alpha$. (c)  $\beta$ versus  $\phi_t$. $\phi_t$ increases $\langle\theta^2\rangle$  and $\beta$. For non of these settings, the conformal behavior emerges.  }
	\label{FIGS3}
\end{figure}

In (2+1)D, as the second scenario, we consider the initial wedge setting as shown in  FIG. \ref{FIGS31} (a).  After running the simulation for 100 time steps,  we analyze the line in $y=0$ plain. Other lines are not statistically identical to this line and should be included. FIGs \ref{FIGS3} (b) and (c) show the effect of $\phi_t$. Fractal dimension analysis for all cases suggests that curves are not conformally invariant.

We summarize results for the effect of the geometry of the environment and initial condition on domain walls in  Table 1.
\begin{table*}  [h]
	\caption{Summarized Main Findings For Domain Walls}
	\setlength{\arrayrulewidth}{0.1mm}
	\setlength{\tabcolsep}{12pt}
	\renewcommand{\arraystretch}{1.5}  
	
	\begin{tabular}{|p{2.9 cm}|p{3.2 cm} |p{3.2 cm} |p{3.2 cm}| }
		
		\hline
		& $\alpha$ & $\langle \theta^2 \rangle $ & $\beta$ \\
		\hline 
		(1+1)D Wedge & Decrease vs  $\phi_l$  &   non-monotonic change &  ---\\ 
		\hline  
		Towards (2+1)D &  Decrease vs $w$  &  Increase vs $w$  & Approaching $\beta=0.32$ \\	
		\hline
		(2+1)D wedge ($\phi_l$)   &  Decrease vs  $\phi_l$ & Decrease vs  $\phi_l$  &    Decrease vs $\phi_l$    \\ 	
		\hline
		(2+1)D wedge ($\phi_t$)   &  Decrease vs  $\phi_t$ & increase vs  $\phi_t$  &    Increase vs  $\phi_t$    \\ 	
		\hline
	\end{tabular}
\end{table*}

\end{document}